\RequirePackage{booktabs}

\documentclass[pdflatex,sn-mathphys-num]{sn-jnl}

\usepackage{graphicx}%
\usepackage{multirow}%
\usepackage{amsmath,amssymb,amsfonts}%
\usepackage{amsthm}%
\usepackage{mathrsfs}%
\usepackage[title]{appendix}%
\usepackage{xcolor}%
\usepackage{textcomp}%
\usepackage{manyfoot}%
\usepackage{booktabs}%
\usepackage{algorithm}%
\usepackage{algorithmicx}%
\usepackage{algpseudocode}%
\usepackage{listings}%

\usepackage{subcaption} 
\usepackage{bbm} 
\usepackage{adjustbox} 
\usepackage{multirow} 
\usepackage{tcolorbox} 
\usepackage{standalone}

\DeclareMathOperator*{\argmax}{arg\,max}
\DeclareMathOperator\erf{erf}
\newcommand{\labelnew}[1]{#1 \colorbox{red!50}{\textcolor{white}{new}}}
\newcommand{\labelmod}[1]{#1 \colorbox{blue!50}{\textcolor{white}{more}}}

\newcommand{\twols}{2LS}
\newcommand{\cbmc}{CBMC}
\newcommand{\cpachecker}{CPAChecker}
\newcommand{\divine}{DIVINE}
\newcommand{\esbmc}{ESBMC}
\newcommand{\pesco}{PeSCo}
\newcommand{\pinaka}{Pinaka}
\newcommand{\uautomizer}{UAutomizer}

\newcommand{\musl}{MUSL}
\newcommand{\coremath}{CORE-MATH}
\newcommand{\bench}[1]{{\em NeuroCodeBench #1}}

\theoremstyle{thmstyleone}%
\theoremstyle{thmstyletwo}%

\theoremstyle{thmstylethree}%

\begin{document}

\title[Floating-Point Neural Network Verification]{Floating-Point Neural Network Verification at the Software Level}


\author*[1]{\fnm{Edoardo} \sur{Manino}}\email{edoardo.manino@manchester.ac.uk}

\author[1]{\fnm{Bruno} \sur{Farias}}\email{bruno.farias@manchester.ac.uk}

\author[1]{\fnm{Rafael S\'a} \sur{Menezes}}\email{rafael.menezes-2@manchester.ac.uk}

\author[2]{\fnm{Fedor} \sur{Shmarov}}\email{fedor.shmarov@newcastle.ac.uk}

\author[1,3]{\fnm{Lucas C.} \sur{Cordeiro}}\email{lucas.cordeiro@manchester.ac.uk}

\affil*[1]{\orgdiv{Department of Computer Science}, \orgname{The University of Manchester}, \orgaddress{\street{Oxford Road}, \city{Manchester}, \postcode{M13 9PL}, \country{UK}}}

\affil[2]{\orgdiv{School of Computing}, \orgname{Newcastle University}, \orgaddress{\street{1 Science Square}, \city{Newcastle upon Tyne}, \postcode{NE4 5TG}, \country{UK}}}

\affil[3]{\orgdiv{Faculty of Electrical and Computer Engineering}, \orgname{Universidade Federal do Amazonas}, \orgaddress{\street{Avenida Rodrigo Otávio}, \city{Manaus}, \postcode{69080-900}, \country{Brazil}}}


\abstract{
The software implementation of neural networks must be proven correct before deployment in safety-critical systems.
To this end, software verification may seem an attractive solution, as automated verification tools can reason about the behaviour of floating-point arithmetic and other typical low-level implementation details of neural network code.
In this paper, we conduct the first rigorous evaluation of eight state-of-the-art automated software verifiers on neural network code.
In doing so, we construct \bench{2.0}, a benchmark comprising $912$ neural network verification examples that cover activation functions, common layers, and full neural networks of up to $170$K parameters.
We demonstrate that the obtained results are rather underwhelming: existing verifiers often return incorrect results, cannot reliably check more than a single neural layer at a time, and do not fully support the standard math library.
To compensate for the latter, we supply the verifiers with explicit implementations of required functions (e.g. TanH), which leads to no visible improvements.
Overall, our study demonstrates that the existing state-of-the-art software verifiers are not ready to handle verification of neural networks at software level.
However, a historical analysis of the progress of verifiers over the years show non-monotonic but constant improvements.
}




\maketitle

\section{Introduction}
\label{sec:intro}

In the past years, there have been growing concerns on the reliability of AI systems and the associated risks~\cite{bengio2025report}. Indeed, AI systems are known to perform exceptionally well on existing benchmarks while still producing an unacceptable rate of incorrect and unsafe predictions in real-world settings~\cite{raji2022fallacy,martinez2024barexam}. Moreover, targeted adversarial attacks are known to arbitrarily degrade the performance of AI systems at inference time~\cite{Szegedy2014intriguing}. While inventing better techniques to benchmark, test~\cite{casper2023redteam}, and improve the reliability of AI systems is an active area of research, an overarching solution is still missing~\cite{Huang2020survey,yang2024survey,PerezCerrolaza2024survey}.

In this regard, neural network verification is emerging as one of the most principled approaches to assess the AI systems' reliability, as it can provide formal guarantees on the behaviour of neural networks~\cite{katz2017reluplex}. In a nutshell, the verification approach takes a neural network model $f$, a safety property $P$ in a suitable formal language, and automatically verifies whether $f$ satisfies $P$ for all possible inputs. For some especially desirable properties, such as robustness to adversarial perturbations of the input~\cite{Szegedy2014intriguing}, state-of-the-art verifiers can scale to neural networks with more than 1M neurons~\cite{brix2024fifth}. Furthermore, research on extending these verification tools to hyper-properties~\cite{teuber2021equivalence,athavale2024hyperproperties}, language models~\cite{Shi2020Robustness,casadio2025antonio}, and cyber-physical systems~\cite{ivanov2020verisig,huang2022polar} is well under way.

At the same time, idealised neural networks models and common neural exchange formats hide most of the implementation details of the final AI system, including floating-point types and numerical optimisations~\cite{Schloegl2023numerical}. Such a level of abstraction is not acceptable in safety-critical applications (CPS, IoT), where guarantees of the implementation's correctness are needed~\cite{abate2017synthesis}. Indeed, type conversion errors in software implementation have caused real-world accidents in the past, ranging from the loss of a \$370 million US dollar space rocket in flight~\cite{lions1996flight} to the death of military personnel due to a faulty air-defence missile~\cite{usgao1992patriot}.

\begin{figure}[t]
\centering
    \includegraphics[width=0.75\textwidth]{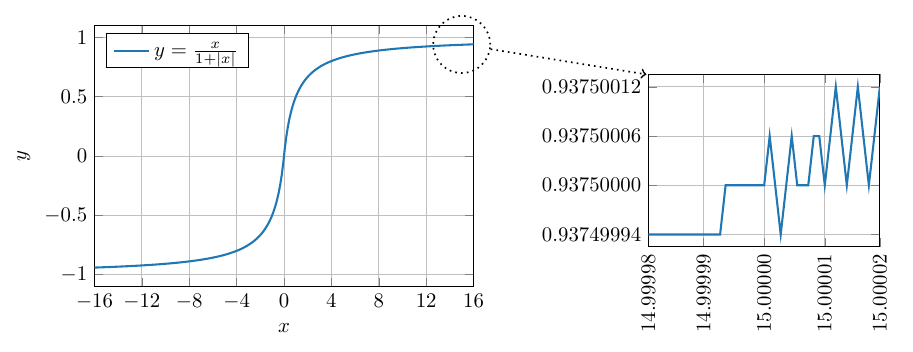}
\caption{The SoftSign activation is a non-decreasing function in infinite precision (left) but not when implemented in 32-bit IEEE 754 floating-point arithmetic (right).}
\label{fig:numerical_issues}
\end{figure}

A simple example of the numerical issues that may arise in floating-point computation is presented in Figure~\ref{fig:numerical_issues}. There, we can see that the SoftSign activation, a common replacement for more computationally expensive sigmoid functions, ceases to be a non-decreasing function once it is implemented in 32-bit IEEE 754 floating point. Establishing whether small numerical errors can cascade into larger vulnerabilities requires reasoning about the software implementation of the entire neural network~\cite{Cordeiro2025esop}. In contrast, existing real-valued verification workflows will hide the issue until a numerically-aware adversary exploits it~\cite{Jia2021floating,Szasz2025soundness}.

In principle, software verification can ensure a system's safety at the implementation level. In fact, verification tools are routinely employed to identify bugs in industrial-scale code or prove their absence~\cite{chudnov2018amazon,fox2023arm}. Unfortunately, existing attempts to employ software verifiers on neural network code have yielded mixed results. On the one hand, software verifiers do not scale well to large neural network instances, forcing practitioners to switch to unsound infinite-precision models~\cite{matos2024ceg4n} or abandoning verification in favour of pure testing approaches~\cite{magalhaes2023c2taco}. On the other hand, software verifiers appear to produce a high rate of incorrect results in some settings~\cite{manino2023neurocodebench}, casting doubt on their role as a safety oracle. Although the latter is a known phenomenon in general software verification tasks~\cite{cadar2016analysing,zhang2019bugs,dyck2023testing}, more work is needed to establish the current capabilities of software verifiers for neural network code and to steer the verification community towards a clear path for improvement.

\textcolor{black}{In this paper, we present the first rigorous study on the verification of floating-point neural network code with automated software verification tools. Our main goal is to validate the claim that these tools support the verification of floating-point code~\cite{beyer2024state}, and reproduce earlier works that suggest that this claim does not extend to neural network implementations~\cite{magalhaes2023c2taco,manino2023neurocodebench}.} To this end, we make the following contributions to the state of the art:
\begin{itemize}
    \item We construct the \bench{2.0} benchmark to be a common ground for comparison across tools. \bench{2.0} contains $912$ snippets of code of varying size: from individual activation functions and neural layers to whole end-to-end neural networks. Furthermore, we augment each benchmark instance with safety specifications in the form of reachability properties; we make sure that the expected verdicts (safe or unsafe) are known in advance via a combination of techniques (e.g., by construction, exhaustive testing). \bench{2.0} is a substantial improvement over our previous short report~\cite{manino2023neurocodebench}, we list all the differences in Appendix \ref{sec:version_diff}.
    \item We make \bench{2.0} compatible with most verification tools by producing fully contained C files that include the model implementation, safety property, required libraries, and system dependencies. In this way, we can rely on the same infrastructure as the International Competition on Software Verification (SV-COMP) ~ \cite{beyer2024state} and guarantee a highly reliable, reproducible workflow via the BenchExec framework~\cite{beyer2019reliable}. \textcolor{black}{Furthermore, our efforts led to \bench{2.0} becoming part of the official benchmark set of SV-COMP since the 2026 edition.}
    \item We compare the performance of eight state-of-the-art automated software verification tools from the perspective of an expert user. Accordingly, we take great care to run the tools with the same settings their authors chose for the 2024 edition of SV-COMP. Overall, we demonstrate that the tools exhibit a large variance across the \bench{2.0} benchmark, with an average of $11\%$ correct and $3\%$ incorrect verdicts.  When we try to improve their behaviour by providing operational models of the C mathematical library in a separate experiment, we show that the performance improvements are minor.
    \item We show how the release of \bench{1.0}, an earlier version of our benchmark, has had a dramatically positive impact on the performance of most tools between 2023 and 2024. \textcolor{black}{At the same time, the introduction of \bench{2.0} in late 2025 demonstrated a more moderate effect on the state-of-the-art software verifiers.} Moreover, we investigate the historical progress of the field by comparing earlier versions of one of the tools (i.e., \esbmc{}) since 2018; our experiments demonstrate a non-monotonic pattern in its performance over the years.
\end{itemize}

In more detail, this paper is structured as follows. In Section \ref{sec:back}, we introduce neural network verification more formally and list the subtle differences introduced by software implementations, floating-point types, and mathematical libraries. In Section \ref{sec:bench}, we present every step of our benchmark design, from initial requirements to the construction of each of its nine benchmark categories. In Section \ref{sec:format}, we discuss our process to make \bench{2.0} compatible with the SV-COMP infrastructure. In Section \ref{sec:exp}, we present the results of our experiments, which include running all verification tools on the plain \bench{2.0} benchmark and adding the operational models of the \texttt{math.h} library, and evaluating the historical progress of the tools. In Section \ref{sec:future}, we draw our conclusions and propose a few suggestions on how to improve the effectiveness of automated software verifiers for the analysis of neural network code.

\section{Background}
\label{sec:back}

\subsection{Neural Networks}
\label{sec:nn_basics}

In machine learning, a neural network is defined as a function $f:\mathbbm{R}^{n_0}\to\mathbbm{R}^{n_L}$ between two real vector spaces of dimensions $n_0$ and $n_L$ respectively~\cite{Bishop2006}. Crucially, modern neural network architectures construct $f$ by composing together a sequence of $L$ simpler functions $\ell_i:\mathbbm{R}^{n_{i-1}}\to\mathbbm{R}^{n_i}$, called layers:
\begin{equation}
    f=\ell_1\circ \ell_2\circ\dots\circ \ell_{L-1}\circ \ell_L
\end{equation}
Common layers include affine transformations such as $\ell_i(x)=Ax+b$ or element-wise activation functions such as $\big(\ell_i(x)\big)_j=\tanh{x_j}$. We give a more comprehensive list of neural network layers and activation functions in Sections~\ref{sec:bench_activ} and~\ref{sec:bench_layers}.

Typically, we optimise the values of layer parameters against a large training set of input-output pairs, until the network reaches the desired accuracy. In some cases, the choice of layer types, layer dimensions $n_i$, and activation functions themselves is the result of an automated process called neural architecture search~\cite{elsken2019nas}. These optimisation processes are governed by a complex set of heuristics, whose hyper-parameters are tuned by trial-and-error to maximise the efficiency of training~\cite{bardenet2013hyperparameters}. In this paper, we assume that training has already finished and we look at the behaviour of the trained neural network $f$.

\subsection{Neural Network Verification}
\label{sec:nn_verification}

While the machine learning process is undoubtedly effective, with state-of-the-art neural networks achieving better than human accuracy in many tasks, it offers no formal guarantees on the correctness of the resulting behaviour. Neural network verification aims to rectify the problem by reasoning over a trained network to prove whether it satisfies certain functional properties~\cite{katz2017reluplex}. More specifically, safety properties over neural networks are usually defined in the following form:
\begin{equation}
\label{eq:safety_prop}
    x\in\mathcal{D}_{in}\implies f(x)\in\mathcal{D}_{out}
\end{equation}
\noindent where $x\in\mathcal{D}_{in}$ is the pre-condition and $f(x)\in\mathcal{D}_{out}$ is the post-condition in computer programming parlance. Perhaps the most famous example of safety property is \textit{adversarial robustness}~\cite{Szegedy2014intriguing}, which requires a neural network to compute the same output class for all possible perturbations of magnitude $\epsilon$ around a given input $c$:
\begin{equation}
\label{eq:adv_robustness}
    ||x-c||\leq\epsilon\implies \argmax\big\{f(x)\big\}=\argmax\big\{f(c)\big\}
\end{equation}
\noindent where the $\argmax$ operator selects the highest scoring out of $n_L$ output classes. In \bench{}, all safety properties are in the form of Equation~\ref{eq:safety_prop}.

Given a neural network $f$ and a safety property $P$, it remains to check whether $P$ holds for $f$. Unfortunately, this decision problem is quite hard to solve. For the most popular networks architectures $f$ and safety properties $P$, complete decision procedures fall into the NP-complete complexity class~\cite{katz2017reluplex}. Furthermore, the worst-case theoretical complexity depends on the number of non-linearities in the network $f$, which correlates with the dimension of the activation layers. In practice, scaling to neural networks containing more than a few million neurons remains a challenge~\cite{brix2024fifth}.

State-of-the-art verification tools use several strategies to solve the aforementioned decision problem. They include convex relaxations of the associated satisfiability problem~\cite{singh2018robustness}, custom SMT solvers~\cite{wu2024marabou}, and over-approximation via abstract domains~\cite{bak2021nnenum}. Currently, the most effective approaches combine branch and bound search with optimisable linear abstraction bounds~\cite{xu2020automatic,xu2021fast}. Since 2020, the international Verification of Neural Networks Competition (VNN-COMP) provides a forum for tool authors to compare progress~\cite{brix2024fifth}. \bench{} contains two categories derived from VNN-COMP benchmarks (see Section \ref{sec:bench_vnn_comp}).

\subsection{Floating-Point Arithmetic}
\label{sec:float_arithmetic}

Although most of the machine learning literature defines neural networks as functions over real vector spaces $f:\mathbbm{R}^{n_0}\to\mathbbm{R}^{n_L}$, the actual computation is run on digital hardware, which can only support finite-precision arithmetic. In this respect, the default numerical type of mainstream machine learning libraries such as Pytorch\footnote{\url{https://pytorch.org}} and Tensorflow\footnote{\url{https://www.tensorflow.org}} is typically the $32$-bit IEEE 754 floating-point format~\cite{IEEE754}. Furthermore, specialised hardware accelerators may favour even shorter types ranging from Google's $16$-bit BFloat16~\cite{bfloat16} to Nvidia's, ARM's and Intel's $8$-bit FP8 format~\cite{Micikevicius2022fp8}.

Compared to real-valued arithmetic, floating-point types introduce a small amount of rounding error after each operation. As a direct consequence, basic arithmetic operations in floating point are not associative~\cite{Shanmugavelu2024impacts}, i.e., we have that $(a+b)+c\neq a+(b+c)$. When executing computationally intensive algorithms, such as convolutions via the Winograd~\cite{Lavin2016winograd} or Fast Fourier Transforms~\cite{pratt2017fouriercnn}, the floating-point deviations compound and lead to different outputs depending on the specific numerical optimisations used~\cite{Schloegl2023numerical}. In neural networks, these numerical discrepancies are large enough to invalidate many of the safety proofs produced by the neural network verifiers we mention in Section~\ref{sec:nn_verification}~\cite{Jia2021floating,Szasz2025soundness}.

As a result, there have been calls to verify neural networks at the implementation level~\cite{Cordeiro2025esop}. That way, we can ensure that the required safety properties hold for the digital artefact $\hat{f}:\mathbbm{F}^{n_0}\to\mathbbm{F}^{n_L}$ that is executed, rather than its real-valued counterpart $f:\mathbbm{R}^{n_0}\to\mathbbm{R}^{n_L}$. Unfortunately, neural network exchange formats such as ONNX\footnote{\url{https://onnx.ai/}} do not contain enough implementation details to aid verification at this level of abstraction~\cite{Jajal2024onnxfailure}. Instead, the few existing studies attempt to specify the details of the neural network implementation at the software level, usually in the C programming language~\cite{manino2023neurocodebench,magalhaes2023c2taco,matos2024ceg4n}. This process essentially turns neural network verification into a software verification problem.

\subsection{Software Verification}
\label{sec:software_verification}

From a high-level perspective, software verification (SV) solves a problem similar to that of neural network verification (VNN) described in Section \ref{sec:nn_verification}. That is, given a program $S$ and a safety property $P$, a software verifier tries to decide whether $P$ holds for all inputs to $S$ (i.e., $S$ is safe with respect to $P$) or not.\footnote{Note that software verification is undecidable in general ~\cite{rice1953classes}. However, we ensure that our benchmarks are always decidable.} If the latter is the case, then the verifier provides a counter-example: a series of input values and assignments that lead to a bug (i.e., safety property violation) in the given program $S$. However, there are two crucial differences between SV and VNN:
\begin{itemize}
    \item Software verifiers need to reason over finite domains (i.e., \texttt{float}'s, \texttt{int}'s, etc.), and therefore, must adhere to the semantics of arithmetic operations over the corresponding domains.
    \item Software verifiers need to make necessary assumptions about the underlying memory model (i.e., how the memory is allocated, accessed, and freed) and computational models of external libraries (e.g., implementations of \texttt{stdlib.h}, \texttt{math.h}, etc.).
\end{itemize}

\begin{figure}[t]%
\centering
    \includegraphics[width=\textwidth]{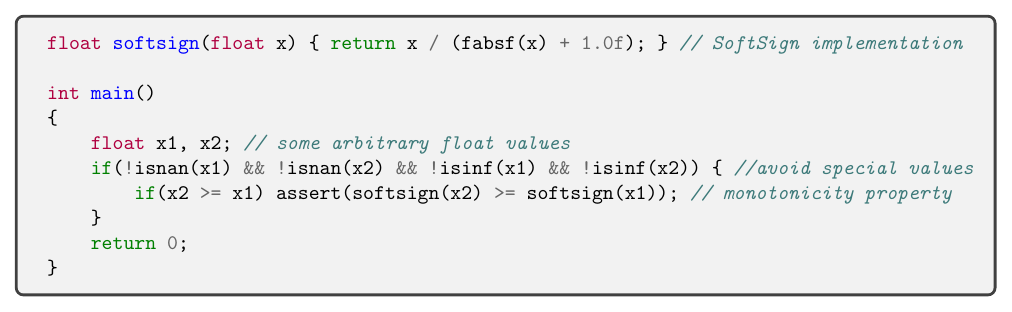}
\caption{Example implementation of SoftSign function from Figure \ref{fig:numerical_issues} together with the monotonicity property that should hold for all values of \texttt{x1} and \texttt{x2} within the numerical limits of the \texttt{float} type. The code is adapted from the file "\texttt{softsign\_2\_unsafe.c}" of \bench{2.0}.}%
\label{fig:sv-example}%
\end{figure}%

In verifying C programs, safety properties are often encoded as \textit{assertions}. Figure~\ref{fig:sv-example} contains a code listing with an implementation of the SoftSign function and the monotonicity property $\forall x_1, x_2 : x_2 \ge x_1 \Rightarrow SoftSign(x_2) \ge SoftSign(x_1)$. As Figure~\ref{fig:numerical_issues} illustrates, the property holds for all $x_1, x_2 \in \mathbb{R}$ but not when $x_1, x_2 \in\mathbbm{F}$. As a consequence, verification of this program will result in an assertion violation. A possible counter-example is \texttt{x1 = 15.0f} and \texttt{x2 = 15.00000286102294921875f}, since \texttt{softsign(15.0f) = 0.9375} and \texttt{softsign(15.00000286102294921875f) = 0.937499940395355224609375}.

There are a myriad of different software verification techniques. Brain et. al. ~\cite{brain2025pyramid} identify six main paradigms: static analysis, abstract interpretation, testing \& symbolic execution, model checking, deductive verification, and functional verification. Many state-of-the-art automated software verifiers (e.g., \cbmc{}~\cite{Kroening2014}, \esbmc{}~\cite{Menezes2024esbmc}, \cpachecker{}~\cite{baier2024}, KLEE~\cite{klee}), cleverly combine different verification techniques across these paradigms. Dozens of such tools participate in SV-COMP~\cite{beyer2024state} -- the largest annual software verification competition -- where they compete in finding bugs (i.e., user-specified assertion violations, buffer overflows, arithmetic overflows, pointer bugs, etc.) in tens of thousands of C programs ranging from simple unit tests to device drivers.

For neural network verification, working at the software level introduces additional scalability issues. The comparison between neural network verifiers and software verifiers in~\cite{matos2024ceg4n} clearly shows a higher timeout rate for the latter. Moreover, the authors of~\cite{magalhaes2023c2taco} observe that software verifiers cannot prove the equivalence of different matrix multiplication algorithms beyond three-dimensional input tensors of size $10\times10\times10$. Our results in Section \ref{sec:results_complexity} confirm that software verifiers struggle with larger benchmark instances.

\subsection{Correctly-Rounded Mathematical Libraries}
\label{sec:op_models}

While software verifiers allow us to reason about floating-point programs, they still rely on assumptions about the semantics of external libraries. In this respect, neural networks often depend on elementary mathematical functions in their implementation of activation functions and network layers, as we show in Sections \ref{sec:bench_maths} and \ref{sec:bench_activ}. For C programs, the function definitions are provided by the standard library \texttt{math.h}, but their behaviour is implementation dependent (see Appendix \ref{sec:ieee754_vs_c}). This lack of portability can become an issue for software verifiers, as they typically need a non-ambiguous specification to correctly reason about the program safety~\cite{Brisebarre2024correctly}.

A principled solution to the ambiguity in floating-point semantics is to provide a reference implementation of the functions in \texttt{math.h}. In this regard, there is a long history of designing explicit approximation algorithms for elementary functions, including CORDIC~\cite{Volder1959cordic}, BKM~\cite{Bajard1994bkm}, and the onion peeling strategy~\cite{Ziv1991rounding}. Over the years, these algorithms have been integrated in the standard mathematical library in conjunction with dedicated hardware primitives for efficiency reasons.\footnote{For an example, see NewLib at \url{https://sourceware.org/newlib/}} In this paper, we consider the following two state-of-the-art implementations of \texttt{math.h}:
\begin{itemize}
    \item \textit{\musl{}}. This mathematical library is based on the well-established numerical algorithms of FDLIBM~\cite{Beebe2017}. While the functions it provides are not always correctly rounded, the documentation\footnote{\url{https://wiki.musl-libc.org/mathematical-library.html}} states that \texttt{sqrtf} is guaranteed to give correctly rounded results and that ``other functions should have small ulp error in nearest rounding mode and the sign of zero should be correct''. Furthermore, the \musl{} library is already used by the software verifier ESBMC for some of its \texttt{math.h} operational models~\cite{Menezes2024esbmc}.
    \item \textit{\coremath{}}. This mathematical library is an evolution of CR-LIBM~\cite{Daramy-Loirat2005crlibm}, a long-standing effort to provide correctly rounded functions~\cite{Sibidanov2022core-math}. Crucially, the numerical algorithms it uses have been formally verified~\cite{deDinechin2011gappa,Martin-Dorel2015formally}. Unfortunately, the library does not provide a square root function, and some trigonometric functions (\texttt{cosf}, \texttt{sinf}) use 128-bit arithmetic, which most software verifiers do not support.\footnote{See code at \url{https://core-math.gitlabpages.inria.fr/}}
\end{itemize}
In Section \ref{sec:results_maths}, we include the MUSL and CORE-MATH numerical algorithms in our benchmark instances, and measure their impact on the performance of existing software verifiers.

\section{Benchmark Design}
\label{sec:bench}

In this section, we present \bench{2.0}, our benchmark of neural network code for software verification. First, we list the design requirements that guided our benchmark-building process. Then we provide details on how we built each benchmark category. Table~\ref{tab:bench_overview} provides an overview of the structure of \bench{2.0}.

\begin{table}[h]
\caption{Overview of \bench{2.0}. The ``Unsafe'' column comprises all properties for which a counterexample exists. The ``Ground Truth'' column reports the source of our verdicts. We highlight differences with \bench{1.0} as \labelnew{} for new categories, and \labelmod{} for expanded ones.}
\label{tab:bench_overview}
\begin{tabular}{llll}
    \toprule
    Benchmark Category & Safe & Unsafe & Ground Truth \\
    \midrule
    \labelmod{Maths Functions} & 42 & 16 & Brute Force \\
    \labelmod{Activation Functions} & 44 & 13 & Brute Force \\
    \labelnew{Neural Layers} & 43 & 43 & By Construction \\
    Hopfield Networks & 46 & 34 & By Construction \\
    \labelnew{SAT ReLU Networks} & 48 & 48 & By Construction \\
    Polynomial Approximation & 48 & 48 & Brute Force \\
    \labelnew{Lipschitz-Bounded Networks} & 54 & 54 & By Construction \\
    Probability Density & 22 & 13 & Counterexamples \\
    Reinforcement Learning & 103 & 193 & Counterexamples \\
    \midrule
    Total & 450 & 462 & \\
    \bottomrule
\end{tabular}
\end{table}

\subsection{Design Requirements}
\label{sec:design}

We design \bench{} with two broad goals in mind: to create a rigorous measurement tool and to stimulate further research on the topic. On the one hand, previous research shows that software verifiers struggle to reason on neural network code~\cite{magalhaes2023c2taco} and sometimes produce incorrect results~\cite{manino2023neurocodebench}. As such, having access to an extensive benchmark of neural network code allows a fair assessment of the performance of state-of-the-art software verification tools. On the other hand, there exist a considerable scalability gap between bit-precise software verification and infinite-precision neural network verifiers~\cite{matos2024ceg4n}. While there are indications that the latter might be solving a fundamentally easier computational problem~\cite{henzinger2021quantized}, further research on reducing the empirical gap between the two requires practical means to measure progress.

Against this background, we design \bench{} with the following three features in mind:
\begin{itemize}
    \item \textit{Ground Truth.} Software verification benchmarks contain assertions over program variables (see Section \ref{sec:software_verification}). In \bench{}, we know \textit{a priori} which assertions hold and which do not, thus making the program safe and unsafe, respectively. Furthermore, we maintain a balance between safe and unsafe verdicts to discourage guessing the answer. As a result, our benchmark can serve as a rigorous measurement tool for the correctness of existing software verifiers. In the remainder of Section \ref{sec:bench}, we give details on the specific techniques we use to establish the ground-truth verdicts.
    \item \textit{Complexity Range.} The scalability of software verifiers is significantly worse than their infinite-precision counterparts~\cite{matos2024ceg4n}. To stimulate further research, we include a wide range of program instances in \bench{}, ranging from very simple mathematical functions to full neural networks. In this way, any improvement in software verification techniques, however small, has a chance to solve a noticeable number of additional instances. We have already seen the positive effects of this incremental reward structure during the 2023 edition of SV-COMP, see Section \ref{sec:results_progress} for details.
    \item \textit{Programming Language.} Mainstream machine learning libraries such as PyTorch\footnote{\url{https://pytorch.org/}} and Tensorflow\footnote{\url{https://www.tensorflow.org}} exhibit inconsistent behaviour across computing platforms~\cite{Schloegl2023numerical}. To avoid the issue, we adopt two microcontroller libraries which convert high-level neural network specifications to standalone artifacts in plain C: onnx2c\footnote{\url{https://github.com/kraiskil/onnx2c}} and keras2c\footnote{\url{https://github.com/PlasmaControl/keras2c}}. Producing benchmarks in plain C has the additional benefit of being compatible with a large number of existing software verifiers~\cite{beyer2024state}. Furthermore, microcontroller platforms are the natural target for a wide range of TinyML and IoT applications, where small neural network models are the norm~\cite{banbury2021micronets,Saha2022microcontroller}.
\end{itemize}

\subsection{Maths Library}
\label{sec:bench_maths}

Typically, neural network code relies on $32$-bit floating-point arithmetic (see Section \ref{sec:float_arithmetic}). In many cases, the code invokes the standard library \texttt{math.h} to perform specific mathematical operations. In particular, this is true for a majority of the activation functions used in neural networks (see Section~\ref{sec:bench_activ}), which are defined in terms of exponential, logarithm, and hyperbolic tangent. Similarly, computing Euclidean distances and vector normalisation require taking the square root~\cite{Bishop2006}. Furthermore, the positional encodings used in some sequence models~\cite{Likhomanenko2021} and convolution algorithms based on the Fourier transform~\cite{pratt2017fouriercnn} depend on the correct implementation of the sine and cosine functions.

\begin{table}[t]
\caption{Safety properties of mathematical functions. Only $42$ out of $58$ instances hold.}
\label{tab:prop_math}
\begin{tabular}{l|rrrrrrrr}
    \toprule
    Properties & \rotatebox{90}{\texttt{cosf}} & \rotatebox{90}{\texttt{erff}} & \rotatebox{90}{\texttt{expf}} & \rotatebox{90}{\labelnew{\texttt{expm1f}}} & \rotatebox{90}{\texttt{logf}} & \rotatebox{90}{\labelnew{\texttt{log1pf}}} & \rotatebox{90}{\texttt{sinf}} & \rotatebox{90}{\texttt{sqrtf}} \\
    \midrule
    Domain            & - & - & - & - & 1 & 1 & - & 1 \\
    Linear Bound      & 2 & 2 & 2 & 2 & 1 & 1 & 2 & 2 \\
    Monotonicity      & - & 1 & 1 & 1 & 1 & 1 & - & 1 \\
    Periodicity       & 1 & - & - & - & - & - & 1 & - \\
    Finite Difference & 2 & 2 & 2 & 2 & 2 & 2 & 2 & 2 \\
    Inverse Function  & 2 & - & 2 & 2 & 2 & 2 & 2 & 2 \\
    Symmetry          & 1 & 1 & - & - & - & - & 1 & - \\
    \midrule
    Total             & 8 & 6 & 7 & 7 & 7 & 7 & 8 & 8 \\
    \bottomrule
\end{tabular}
\end{table}

In this light, we provide $58$ benchmark instances that check whether a software verifier can reason over the mathematical functions in Table~\ref{tab:prop_math}. In addition to the functions listed there, our benchmark relies on the following ancillary functions: \texttt{acosf}, \texttt{asinf}, \texttt{atanhf}, \texttt{fabsf}, \texttt{fmaxf}, \texttt{isgreater}, \texttt{isgreaterequal}, \texttt{isless}, \texttt{islessequal}. Furthermore, we include the function \texttt{tanhf} under the category of activation functions (see Section \ref{sec:bench_activ}). We defer a discussion on the inherent ambiguity of writing bit-precise specifications over \texttt{math.h} functions to Section \ref{sec:op_models}.

Overall, our benchmark checks whether the floating-point implementations of the functions in Table \ref{tab:prop_math} satisfy the fundamental properties of their corresponding mathematical functions. More specifically, we check for the following properties:
\begin{itemize}
    \item \textit{Domain.} Logarithms and square root return \texttt{NaN} for negative inputs.
    \item \textit{Linear Bound.} All functions satisfy some linear bounds in the form $\pm f(x)\leq ax+b$. For instance, $\cos(x)\leq1$ and $exp(x)\geq x+1$.
    \item \textit{Monotonicity.} Most functions are monotonically non-decreasing.
    \item \textit{Periodicity.} Sine and cosine have a period of $2\pi$.
    \item \textit{Finite Difference.} All functions satisfy some invariants on their derivatives $f'(x)$. For instance, $\erf'(x)\leq 2/\sqrt{\pi}$ and $\exp'(x)=\exp(x)$.
    \item \textit{Inverse Function.} Most functions have a closed-form inverse. For instance, $\arccos(\cos(x))=x$ and $\log(\exp(x))=x$.
    \item \textit{Symmetry.} The error function, sine, and cosine are symmetric around zero. For instance, $\erf(x)=-\erf(-x)$ and $\cos(x)=\cos(-x)$.
\end{itemize}
Due to the non-idealities of floating-point representation, many of the mathematical properties listed above do not always hold (e.g., see Figure~\ref{fig:numerical_issues}). For our benchmark, we derived the ground-truth verdicts in Table \ref{tab:bench_overview} by checking each safety property against all possible $32$-bit floating-point inputs.

\subsection{Activation Functions}
\label{sec:bench_activ}

\begin{figure}[t]
\centering
    \begin{subfigure}[b]{0.455\textwidth}
        \includegraphics[width=\textwidth]{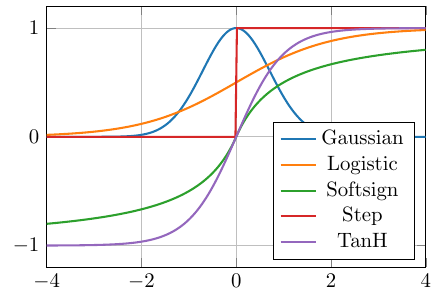}
        \caption{Bounded activations}
        \label{fig:activations_bounded}
    \end{subfigure}
    \begin{subfigure}[b]{0.44\textwidth}
        \includegraphics[width=\textwidth]{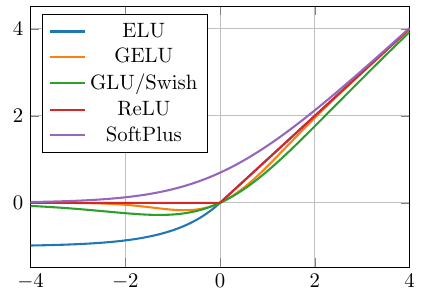}
        \caption{Unbounded activations}
        \label{fig:activations_unbounded}
    \end{subfigure}
\caption{Visual comparison of popular activation functions.}
\label{fig:activations}
\end{figure}

The vast majority of neural network architectures include several activation layers, where the output of each neuron is passed through a so-called activation function (see Section \ref{sec:nn_basics}). These functions are highly non-linear and allow modern neural networks to fit arbitrary datasets~\cite{Bishop2006}. We can classify all activation functions based on the range of their co-domain: bounded or unbounded (see Figure~\ref{fig:activations}). In our benchmark, we include the following popular activation functions~\cite{Nwankpa2021,Hendrycks2023}:
\begin{itemize}
    \item ELU$(x)=\{\exp(x)-1$ if $x<0$, $x$ otherwise$\}$.
    \item Gaussian$(x)=\exp(-x^2)$.
    \item GELU$(x)=x\cdot\frac{1}{2}\big(1+\erf(x/\sqrt{2})\big)$.
    \item GLU$(x,y)=x\cdot$Logistic$(y)$. Also called Swish when $x=y$.
    \item Logistic$(x)=\frac{1}{2}\big(1+\tanh(x/2)\big)=\big(1+\exp(-x)\big)^{-1}$. Also called Sigmoid.
    \item ReLU$(x)=\{0$ if $x<0$, $x$ otherwise$\}=\max(0,x)$.
    \item SoftPlus$(x)=\log(1+\exp(x))$.
    \item SoftSign$(x)=x\cdot(1+|x|)^{-1}$.
    \item Step$(x)=\{0$ if $x<0$, $1$ otherwise$\}$.
    \item TanH$(x)=\big(\exp(2x)-1\big)\cdot\big(\exp(2x)+1\big)^{-1}$.
\end{itemize}

\begin{table}[t]
\caption{Safety properties of common activation functions. Only $44$ of our $57$ instances hold.}
\label{tab:prop_activ}
\begin{tabular}{l|rrrrrrrrrr}
    \toprule
    Properties & \rotatebox{90}{ELU} & \rotatebox{90}{\labelnew{Gaussian}} & \rotatebox{90}{GELU} & \rotatebox{90}{\labelnew{GLU}} & \rotatebox{90}{Logistic} & \rotatebox{90}{ReLU} & \rotatebox{90}{SoftPlus} & \rotatebox{90}{SoftSign} & \rotatebox{90}{\labelnew{Step}} & \rotatebox{90}{TanH} \\
    \midrule
    Linear Bound      & 2 & 2 & 3 & 2 & 2 & 2 & 3 & 2 & 2 & 2 \\
    Monotonicity      & 1 & - & 1 & 2 & 1 & 1 & 1 & 1 & 1 & 1 \\
    Finite Difference & 2 & 2 & 2 & - & 2 & - & 2 & 2 & - & 2 \\
    Inverse Function  & - & - & - & - & 2 & - & - & - & - & 2 \\
    Symmetry          & - & 1 & - & 2 & 1 & - & - & 1 & 1 & 1 \\
    \midrule
    Total             & 5 & 5 & 6 & 6 & 8 & 3 & 6 & 6 & 4 & 8 \\
    \bottomrule
\end{tabular}
\end{table}

Overall, we write $57$ safety properties of activation functions. In a similar fashion to the mathematical library (see Section \ref{sec:bench_maths}), we exploit several fundamental properties to construct our benchmark. We list them in Table \ref{tab:prop_activ}. Note that, differently from the properties in Table \ref{tab:prop_math}, none of the activation functions we consider is periodic or has a limited input domain.

\subsection{Neural Layers}
\label{sec:bench_layers}

One of the design requirements of \bench{} is that it should contain instances of progressively increasing difficulty (see Section \ref{sec:design}). In this respect, previous work shows that software verifiers struggle to reason about the implementation of simple linear algebra primitives as soon as their dimension exceeds toy examples~\cite{magalhaes2023c2taco}. As neural networks make a heavy use of such primitives, it is worth including them in our benchmark.

More specifically, we focus on the basic constituent layers of most neural networks~\cite{Bishop2006}. Typically, they fall into the following four subcategories:
\begin{itemize}
    \item \textit{Affine Layers.} One of the most computationally intensive operations in neural networks are affine transformations, i.e., they map an input tensor $x$ to an output $y=Ax+b$ \cite{gholami2022survey}. Here, we include a so-called Linear layer,\footnote{\url{https://pytorch.org/docs/stable/generated/torch.nn.Linear.html}} where $Ax$ is a dense matrix-vector multiplication, and a simplified version of a one-dimensional convolutional layer,\footnote{\url{https://pytorch.org/docs/stable/generated/torch.nn.Conv1d.html}} where $A$ is sparse.
    \item \textit{Normalisation Layers.} The learning dynamics of gradient descent can be greatly improved by normalising the tensors in the hidden layers of a neural network~\cite{santukar2018optimisation}. Batch normalisation~\cite{ioffe2015batchnorm} achieves this by rescaling all tensors in a training batch to have zero mean and unitary standard deviation. Layer normalisation~\cite{ba2016layer} achieves the same result by computing the mean and standard deviation across an individual tensor, rather than a whole batch. Both normalisation layers rely on computing a square root.
    \item \textit{Pooling Layers.} Mostly used in convolutional neural networks, pooling layers downsample an input tensor to a lower-dimensional one~\cite{scherer2010pooling}. The downsampling is controlled by setting a window size $w$ and stride $s$. For one-dimensional tensors, pooling is defined as $y_i=$ Pool$(x_{(i-1)s},\dots,x_{(i-1)s+w-1})$, where MaxPool and AvgPool return the maximum and average value, respectively.
    \item \textit{SoftMax Layer.} Defined as SoftMax$(x)=\exp(x)/\sum_i^d\exp(x_i)$,\footnote{For simplicity, we omit the temperature $\tau$ from our implementation.} it takes an arbitrary input vector $x\in\mathbbm{F}^d$ and maps it to a valid probability distribution~\cite{Bishop2006}. The presence of element-wise exponentials makes it challenging for software verifiers.
\end{itemize}

As Table \ref{tab:prop_layer} shows, we write $86$ safety properties across the seven neural layers. We keep the layer size small ($d\in[2,10]$) to provide a moderate increase in verification complexity with respect to the individual functions in Sections \ref{sec:bench_maths} and \ref{sec:bench_activ}. In contrast with the categories therein, we take advantage of the additional properties of identity ($Ax=x$ when $A=I$ is the identity matrix) and of the fact that the outputs of LayerNorm and SoftMax should have norm one.

\begin{table}[t]
\caption{Safety properties of basic neural layers. Only $43$ out of $86$ instances hold.}
\label{tab:prop_layer}
\begin{tabular}{l|rrrrrrr}
    \toprule
    Properties & \rotatebox{90}{\labelnew{AvgPool}} & \rotatebox{90}{\labelnew{BatchNorm}} & \rotatebox{90}{\labelnew{Conv1D}} & \rotatebox{90}{\labelnew{LayerNorm}} & \rotatebox{90}{\labelnew{Linear}} & \rotatebox{90}{\labelnew{MaxPool}} & \rotatebox{90}{SoftMax} \\
    \midrule
    Linear Bound & 11 &  - &  - &  - &  - & 11 &  6 \\
    Monotonicity &  3 &  4 &  - &  8 &  - &  3 &  4 \\
    Symmetry     &  - &  8 &  4 &  - &  4 &  - &  - \\
    Identity     &  - &  - &  4 &  - &  8 &  - &  - \\
    Output Norm  &  - &  - &  - &  4 &  - &  - &  4 \\
    \midrule
    Total        & 14 & 12 &  8 & 12 & 12 & 14 & 14 \\
    \bottomrule
\end{tabular}
\end{table}

\subsection{Hopfield Networks}
\label{sec:bench_hopfield}

\begin{figure}[t]
\centering
    \includegraphics[width=0.75\textwidth]{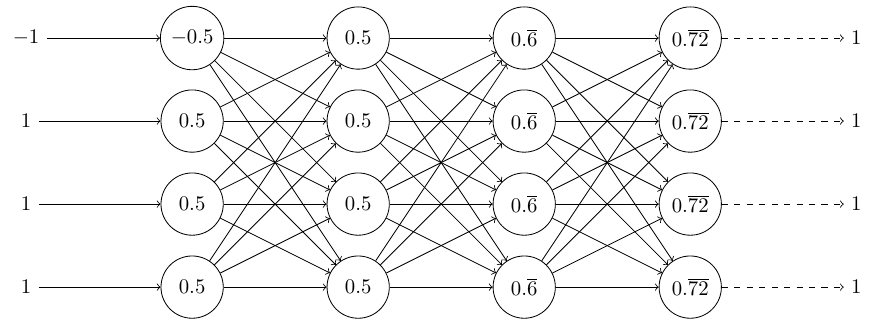}
\caption{An (unrolled) Hopfield network with SoftSign activations and $t=4$ recurrent steps. The network is able to recover the correct (positive) sign for all output entries, when a strict majority of the input entries are correct. We omit the final classification layer (dashed) from our benchmarks.}
\label{fig:hop_w4_softsign}
\end{figure}

Hopfield networks are known for their use as error-correcting decoders~\cite{AbuMostafa1985}. From a practical perspective, the Hopfield neural architecture takes an input vector of $d$ sign bits $x\in\{\pm1\}^d$, and produces an output $y\in\{\pm1\}^d$ of the same type and size. Modern variants of Hopfield networks can memorise a large number of vectors during training, and correct any entries in the input vector that had their sign flipped~\cite{Chaudhuri2019}. We give a visual example of the computation required to do so in Figure \ref{fig:hop_w4_softsign}.

For \bench{}, we adopt a simplified Hopfield architecture based on Hebbian weights, which does not require training~\cite{AbuMostafa1985}. Specifically, we construct our models to reconstruct the individual input vector $x^*=\mathbf{1}$, whose entries are all unitary. Furthermore, we encode our Hopfield networks in the following recurrent form:
\begin{equation}
\label{eq:recurrent_net}
    y_{t+1} = \sigma(Uy_t + Vx_t + b)
\end{equation}
\noindent where the input matrix is the identity $V=I$, the kernel (state) matrix has all unitary entries $U=\mathbf{1}\mathbf{1}^T$, the bias is $b=0$, the initial state is $y_0=0$ and the input is non-zero at the initial step only $x_{t\geq1}=0$. Traditionally, Hopfield networks use the Step activation function $\sigma$. For compatibility with keras2c~\cite{Conlin2021}, we replace it with either the SoftSign or the TanH activation instead. Overall, we construct $40$ different Hopfield network by varying the activation function $\sigma\in\{$SoftSign$,$TanH$\}$, the input vector size $d\in\{4,8,16,32,64\}$ and the number of timesteps $t\in\{1,2,3,4\}$, as shown in Table~\ref{tab:nn_arch}.

Finally, we write $80$ safety properties by exploiting the fact that the network in Equation \ref{eq:recurrent_net} can always reconstruct $y=x^*$, given enough timesteps, when at least $d/2+1$ of the input entries are correct. As such, we always set the first $d/2-1$ entries of $x_0$ to non-deterministic floating-point numbers in $[-1,1]$. Thanks to the monotonicity of the Hopfield model, it is possible to check its worst and best-case behaviour by testing the two concrete inputs $x_0^-=(-1,\dots,-1,1,\dots,1)$ and $x_0^+=(1,\dots,1)$ respectively. An example of such analysis for $d=4$, $t=4$ and $\sigma=$ SoftSign is shown in Figure \ref{fig:hop_w4_softsign}. With the technique of propagating $x_0^-$ and $x_0^+$ through the network $f$, we are able to establish provably-correct post-conditions in the form $y_t\geq f(x_0^-)$ and $y_t\leq f(x_0^+)$.

\subsection{SAT ReLU Networks}
\label{sec:bench_gadgets}

A classic result on the theoretical complexity of neural network verification states that it is as hard as solving a 3-SAT problem~\cite{katz2017reluplex}. In deriving their proof, the authors introduce several gadgets to embed 3-SAT formulae into ReLU neural networks. Here, we exploit and extend their theoretical construction to turn arbitrary SAT formulae into neural network verification benchmarks.

First, let us establish the following gadgets, which allow us to declare Boolean variables over the binary set $\{0,1\}$ and define the semantics of the operators $\lnot$, $\lor$, and $\land$:
\begin{itemize}
    \item \textit{Binarisation.} Given $x\in[0,1]$, we have $x\in\{0,1\}$ iff $x-$ReLU$(2x-1)=0$.\footnote{The authors of~\cite{katz2017reluplex} propose the alternative condition ReLU$(\epsilon-x)+$ReLU$(x-1+\epsilon)\leq\epsilon$ for any arbitrarily small $\epsilon$ (see Appendix I therein). Since said condition is always satisfied for $x\in[0,1]$, it does \textit{not} ensure that $x\in[0,\epsilon]\cup[1-\epsilon,1]$ as the authors claim. Our binarisation condition $x-$ReLU$(2x-1)=0$ solves the issue.}
    \item \textit{Negation.} Given $x,y\in\{0,1\}$, we have $y=\lnot x$ iff $y=1-x$.
    \item \textit{Disjunction.} Given $x_i\in\{0,1\}$, we have $y=x_1\lor \dots\lor x_n$ iff $y=1-\text{ReLU}(1-\sum_{i=1}^nx_i)$.
    \item \textit{Conjunction.} Given $x_i\in\{0,1\}$, we have $y=x_1\land \dots\lor x_n$ iff $y=\text{ReLU}(1-n+\sum_{i=1}^nx_i)$.
\end{itemize}

\begin{figure}[t]
\centering
    \includegraphics[width=0.7\textwidth]{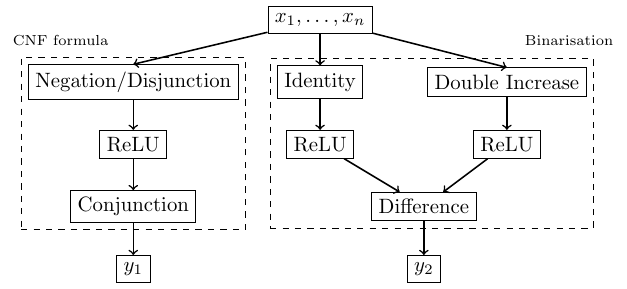}
\caption{Details of the SAT ReLU network architecture (a sketch).}
\label{fig:sat_relu_net}
\end{figure}

Next, we show how to encode an arbitrary SAT formula $P$ in the form of a neural network (see Figure \ref{fig:sat_relu_net} on the left). Without loss of generality, assume that $P$ is in conjunctive normal form (CNF), i.e., $P=\bigwedge_{i}c_i$ where $c_i=\bigvee_jt_j$ and $t_j\in\{x_k,\lnot x_k\}$. Additionally, assume for now that $x_k\in\{0,1\}$ for all $k$. We encode the negation of each disjunctive clause $c_i$ in a separate neuron $\bar{c}_i $ of a fully-connected ReLU layer as follows:
\begin{equation}
\label{eq:sat_disjunction}
    \bar{c} = \sigma(Wx+b)
    \qquad\text{where}\qquad W_{ik}=
    \begin{cases}
        -1 \text{ if } x_k\in c_i\\
        +1 \text{ if } \lnot x_k\in c_i\\
        0 \text{ otherwise}
    \end{cases}
    \qquad\text{and}\qquad b_i=1-\sum_{\lnot x_k\in c_i}1
\end{equation}
\noindent where $\sigma$ is the ReLU activation function and $x_k\in c_i$ (respectively $\lnot x_k\in c_i$) if variable $x_k$ (or its negation) appears as a term of clause $c_i$. Given $\bar{c}$ in Equation \ref{eq:sat_disjunction}, we encode the conjunction of their negated value as $y_1=1-\mathbf{1}^T\bar{c}$, where $\mathbf{1}^T$ is a row vector with unitary entries. Note that we omit the ReLU activation from the conjunction gadget for now.

Similarly, we can extend our neural network encoding to include the binarisation gadgets for the input vector $x$ (see Figure \ref{fig:sat_relu_net} on the right). To do so, we introduce the following fully-connected ReLU layer:
\begin{equation}
\label{eq:sat_binarisation}
    b=\sigma(Vx+u)
    \qquad\text{where}\qquad V=
    \begin{bmatrix}
        I\\
        2I
    \end{bmatrix}
    \qquad\text{and}\qquad u=
    \begin{bmatrix}
        \mathbf{0}\\
        -\mathbf{1}
    \end{bmatrix}
\end{equation}
\noindent where $\sigma$ is the ReLU activation function, $I$ is the identity matrix and $\mathbf{0}$, $\mathbf{1}$ are column vectors with zero-valued and unitary entries, respectively. Note that the presence of the ReLU activation is only necessary for half of the binarisation gadget, but we keep it for the whole layer for reason of simplicity. Given $b$ in Equation \ref{eq:sat_binarisation}, we encode the result of the binarisation gadget as $y_2=[\mathbf{1}^T -\mathbf{1}^T]b$.

Finally, we can express the SAT problem associated to any arbitrary CNF formula $P$ as a safety property over its equivalent ReLU neural network. Specifically, $P$ is satisfiable if and only if there exists some $x\in[0,1]^n$ such that $y_1=1$ and $y_2=0$. The former post-condition ($y_1=1$) takes the place of the ReLU activation function in the conjunction gadget, as any input $x\in\{0,1\}^n$ that does not satisfy $P$ will cause $y_1\leq0$. The latter condition ($y_2=0$) ensures that the inputs have been binarised correctly.

In total, we generate $96$ SAT ReLU benchmarks from boolean formulae of different sizes. We vary the number of input variables in $n\in\{5,8,13,21,34,55,89,144\}$ and the number of clauses in $|c|=nr$ with the multiplier $r\in\{1,2,3,4,5,6\}$. On the one hand, we generate satisfiable formulae by choosing a satisfying assignment, and then creating $|c|$ random clauses that do not contradict it. On the other hand, we generate unsatisfiable formulae iteratively: we begin with the contradiction $x_1\land\lnot x_1$, then we pick an existing clause $c_i$ at random and replace it with the two clauses $c_i\lor x_k$ and $c_i\lor \lnot x_k$, where $x_k\not\in c_i$. The size of the corresponding neural network is $n\times(|c|+2n)\times2$, where $|c|$ hidden neurons are needed for the disjunction gadgets and $2n$ for the binarisation ones (see Figure \ref{fig:sat_relu_net}).

\subsection{Polynomial Approximation Networks}
\label{sec:bench_poly}

In contrast to the classification problems presented in Sections \ref{sec:bench_hopfield} and \ref{sec:bench_gadgets}, neural networks are also routinely employed to approximate the behaviour of several physical and electrical processes by solving the corresponding regression problem (see for instance~\cite{Xu2002,Massi2023}). In such applications, we assume that there is an unknown function $g$ that we wish to approximate with a neural network $f$, given a limited set of samples $\{(x,g(x))_i\}$. Ideally, we would like the approximation $f$ to be as close to the unknown function $g$ as possible.

\begin{figure}[t]
\centering
    \includegraphics[width=0.75\textwidth]{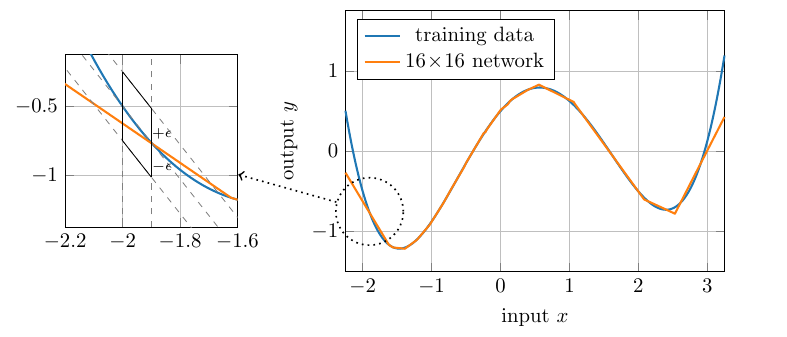}
\caption{Comparison between one of our polynomial approximation networks and its training data (right). The safety property requires that the output error of the network does not exceed $\epsilon$ in a small input interval (left).}
\label{fig:poly_prop}
\end{figure}

Here, we emulate this process for an arbitrary oscillating polynomial $g(x) = 0.125 x^4 - 0.25 x^3 - 0.75 x^2 + x + 0.5$ (see Figure \ref{fig:poly_prop}), from which we extract $100$ points by linearly sampling the interval $x\in[-2,3]$. Then, we train several multilayer perceptrons $f$, whose architecture comprises $L$ layers of the following form:
\begin{equation}
    x_{\ell}=\sigma_{\ell}(W_{\ell}x_{\ell-1}+b_{\ell})
\end{equation}
\noindent where $\ell$ is the layer index, $x_0$ is the network input, $x_L$ is the network output, $W_{\ell}$ is the weight matrix, $b_{\ell}$ is the bias vector, $\sigma_{\ell}$ is the ReLU activation function when $\ell<L$ and the identity function for the last (output) layer. In total, we train $16$ multilayer perceptrons with different number of layers $L\in\{2,3,4,5\}$, as shown in Table~\ref{tab:nn_arch}. For each network, we keep the number of neurons $d$ equal for all its hidden layers $\ell\in[1,L-1]$, such that $x_{\ell}\in\mathbbm{F}^d$. We vary $d\in\{4,8,16,32,64,128,256,512,1024\}$, as long as the total number of neurons does not exceed $256$ when $L>2$. Figure \ref{fig:poly_prop} shows the behaviour of the $L=3$, $d=16$ network after training.

We define our safety properties as follows. First, we formally evaluate the quality of the approximation $f(g)\approx g(x)$ by scanning every possible 32-bit floating-point value in the training interval $x\in[-2,3]\subset\mathbbm{F}$. For each network $f$, we identify the input that causes the largest difference $x^*=\argmax_x|f(x)-g(x)|$, and set our pre-condition to be a small interval around $x^*$ of size $0.1$ (see Figure \ref{fig:poly_prop} on the left). Second, we compute the maximum difference $\epsilon^*=\max|f(x)-\hat{g}(x)|$ against an affine approximation $\hat{g}$ of the polynomial $g$ in the small interval around $x^*$ that satisfies the pre-condition. We use this information to generate six different post-conditions in the form $|f(x)-\hat{g}(x)|\leq\epsilon$ (see Figure \ref{fig:poly_prop} on the left): three of them set $\epsilon<\epsilon^*$ thus yielding a counterexample, three of them set $\epsilon>\epsilon^*$ thus yielding safe instances. We choose the six values of the threshold $\epsilon\in\{\epsilon_1,\dots,\epsilon_6\}$ in small increments such that $\epsilon_i=2\epsilon_{i-1}$. Values of $\epsilon_i$ further away from the satisfiability threshold $\epsilon^*$ usually yield benchmark instances that are easier to solve~\cite{huang2024towards}. In Section \ref{sec:results_complexity_poly_approx}, we show that the latter is not true for software verifiers.

\subsection{Lipschitz-Bounded Networks}
\label{sec:bench_lipschitz}

Given the relative fragility of neural networks to small input perturbations~\cite{Biggio2013evasion}, the machine learning community has dedicated considerable effort to quantifying and controlling their effects~\cite{Tsuzuku2018lipschitz,Cohen2019smoothing}. One popular approach relies on controlling the Lipschitz constant $c$ of the whole network $f$, such that $\lVert f(x)-f(x')\rVert_p\leq c\lVert x-x'\rVert_p$ for any pair of arbitrary inputs $x$, $x'$ in some norm $p$~\cite{Zhang2022lipschitz,Araujo2023sll}. In this way, any small input perturbation $x'=x+\epsilon$ cannot cause a large change in the network output.

Here, we adopt the Lipschitz-bounded network architecture in~\cite{Araujo2023sll}. It is composed by residual layers $\ell$ with the following form:
\begin{equation}
\label{eq:sll_layer}
    x_{\ell}=x_{\ell-1}-2W_{\ell}^TD^{-1}\sigma(W_{\ell}x_{\ell-1}+b_{\ell})
\end{equation}
where $\sigma$ is the ReLU activation function, and $D=$ diag$(\sum_{j=1}^d|W_{\ell}W_{\ell}^T|_{ij}q_j/q_i)$ with a vector $q$ of additional trainable parameters. Note that residual layers like the one in Equation \ref{eq:sll_layer} require the inputs $x_{\ell-1}\in\mathbbm{F}^d$ and outputs $x_{\ell}\in\mathbbm{F}^d$ to have the same dimension $d$. We can increase $d$ by introducing an additional zero-padding layer and decrease $d$ by introducing a (normalised) linear projection.\footnote{See the code at \url{https://github.com/araujoalexandre/Lipschitz-SLL-Networks/tree/main/core/models}} By construction, Equation \ref{eq:sll_layer} has a Lipschitz constant of $c=1$ in Euclidean norm ($p=2$)~\cite{Araujo2023sll}.

\begin{figure}[t]
\centering
    \begin{subfigure}[b]{0.52\textwidth}
        \includegraphics[width=\textwidth]{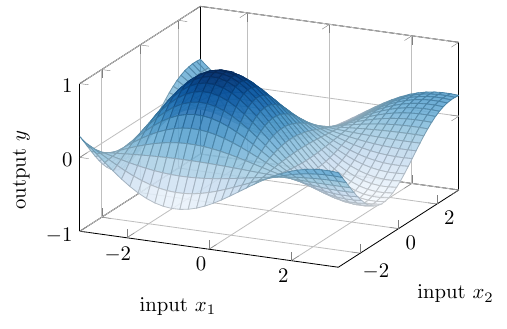}
        \caption{Training data with two input dimensions}
        \label{fig:lipschitz_gt}
    \end{subfigure}
    \hfill
    \begin{subfigure}[b]{0.46\textwidth}
        \includegraphics[width=\textwidth]{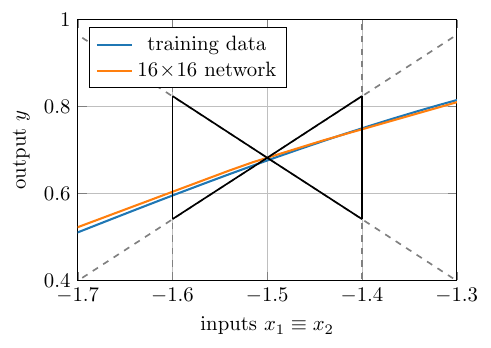}
        \caption{Lipschitz bounds on the $x_1\equiv x_2$ line}
        \label{fig:lipschitz_bounds}
    \end{subfigure}
\caption{Safety properties of Lipschitz-bounded networks.}
\label{fig:lipschitz_prop}
\end{figure}

To construct our benchmark, we train $18$ different Lipschitz-bounded neural networks with $k\in\{2,3,4\}$ inputs and two residual layers of dimension $d\in\{4,8,12,16,20,24\}$. We construct our training set by extracting $65536$ samples from the following oscillating function:
\begin{equation}
\label{eq:lipschitz_gt}
    y=\prod_{i=1}^kg(x_i)\qquad\text{where}\qquad g(x)=0.75\cos\left(0.75 x\right)-0.5\sin(x)
\end{equation}
\noindent where the inputs are taken uniformly at random from the multi-dimensional interval $x\in[-\pi,\pi]^k$. In Figure \ref{fig:lipschitz_gt}, we show a visual representation of Equation \ref{eq:lipschitz_gt} for $k=2$.

Thanks to the property of Lipschitz-boundedness, we can define safety properties over the behaviour of each $f$ without the need of the same computationally-expensive procedure from Section \ref{sec:bench_poly}. More specifically, we restrict our pre-conditions to the input interval $x\in[-1.6,-1.4]^k$, which is centered around a region of maximum slope. Then, we compute the output of the network at the centroid $x_c=(-1.5,\dots,-1.5)$ which automatically yields the conical-shaped output bounds shown in Figure \ref{fig:lipschitz_prop}. In fact, knowing that $c=1$ for $p=2$, we can conclude that $f(x)\in[f(x_c)-0.1\sqrt{k},f(x_c)+0.1\sqrt{k}]$ where $0.1$ is the maximum perturbation around the input $x_c$. With this, we write three post-conditions in the form $f(x)\leq f(x_c)+0.1\sqrt{k}s$ with $s\in\{1,2,4\}$ for each network $f$.

To complete our Lipschitz-bounded benchmark, we add a set of violated properties. We construct these by testing $10$M random inputs and establishing a lower bound on the maximum possible output $\hat{y}\leq\max_xf(x)$ in the pre-condition interval $x\in[-1.6,-1.4]^k$. Then, we write a further three post-conditions in the form $f(x)\leq f(x_c)+(\hat{y}-f(x_c))/s$ with $s\in\{1,2,4\}$ for each network $f$. In total, the Lipschitz-bounded benchmark contains $108$ safety properties (see Table \ref{tab:bench_overview}).

\subsection{VNN-COMP Networks}
\label{sec:bench_vnn_comp}

The International Verification of Neural Network Competition (VNN-COMP) provides an extensive corpus of benchmarks dating back to its inception in 2020~\cite{Mueller2023}. These benchmarks use the Open Neural Network eXchange (ONNX) format\footnote{\url{https://onnx.ai/}} to specify a neural network's parameters and architecture. While most VNN-COMP benchmarks use 32-bit floating point constants, the ONNX format does not necessarily mandate the specific order of arithmetic operations. As a result, executing the same ONNX neural network on different software and hardware platforms may yield inconsistent results~\cite{Schloegl2023numerical}. Indeed, the verifiers that compete in VNN-COMP are supposed to reason on the real-valued behaviour of such neural networks, thus side-stepping the non-idealities of floating-point arithmetic.

Here, we propose a straightforward methodology for translating any VNN-COMP benchmark to a lower-level abstraction defined by C code. We do so in two steps. First, we provide a reference implementation of an ONNX neural network by running it through the onnx2c tool.\footnote{\url{https://github.com/kraiskil/onnx2c}} Second, we convert the corresponding safety properties from the original VNN-LIB format\footnote{\url{https://www.vnnlib.org/}} to a minimal \texttt{main()} function with explicit pre- and post-conditions on the inputs and outputs of the neural network.

More concretely, we apply our translation methodology to the two specific categories listed below, which first appeared in the 2022 edition of VNN-COMP. As opposed to the majority of the benchmarks therein, the two categories below involve neural networks of comparable size to the ones in \bench{} (see Table \ref{tab:nn_arch}):
\begin{itemize}
    \item \textit{Probability Density.} This is a corpus of three neural networks that estimate the probability density of the evolution trajectories in various state system~\cite{Meng2022}: ground collision avoidance systems for F-16 aircrafts (GCAS), ground robot navigation (Robot), Van der Pol oscillators (VDP). The corpus is associated with $37$ safety properties in total, which are derived via reachability techniques.
    \item \textit{Reinforcement Learning.} This is a corpus of three neural networks trained to solve some reinforcement learning tasks with safety constraints~\cite{Ravaioli2022}: the cart-pole toy control problem, Dubins rejoin manoeuvre in aircraft formation flight, space docking manoeuvre for a lunar lander. The corpus is associated with $296$ safety properties corresponding to the safety constraints the neural networks are trained to satisfy.
\end{itemize}

\begin{table}[t]
\caption{Architecture and size of the neural networks in \bench{2.0}.}
\label{tab:nn_arch}
\begin{tabular}{l|lllllll}
    \toprule
    Subcategory & Arch. & In & Out & Depth & Width & Act. Fun. & Convert \\
    \midrule
    \multirow{2}{*}{Hopfield Networks} & \multirow{2}{*}{Recur.} & \multirow{2}{*}{4--64} & \multirow{2}{*}{4--64} & \multirow{2}{*}{1} & \multirow{2}{*}{4--64} & TanH/ & \multirow{2}{*}{keras2c} \\
     &  &  &  &  &  & SoftSign &  \\
    \labelnew{SAT ReLU Networks} & FC & 5--144 & 2 & 1 & 15--1152 & ReLU & onnx2c \\
    Polynomial Approximation & FC & 1 & 1 & 1--4 & 4--1024 & ReLU & keras2c \\
    \labelnew{Lipschitz-Bounded Nets} & Resid. & 2--4 & 1 & 2 & 4--24 & ReLU & onnx2c \\
    Probability Density & FC & 3--14 & 3--14 & 2--3 & 32--64 & ReLU & onnx2c \\
    Reinforcement Learning & FC & 4--8 & 2--8 & 2 & 64--256 & ReLU & onnx2c \\
    \bottomrule
\end{tabular}
\end{table}

A major downside of our translation methodology is that we lose any soundness guarantee on the verdict associated to the safety properties. Indeed, by considering the bit-precise behaviour of a very specific C implementation in 32-bit floating point, we might be introducing or removing vulnerabilities with respect to the real-valued version of the benchmark. While we have no option but to trust the safety verdicts of the real-valued VNN-COMP verifiers, we can at least recover the soundness of the benchmarks on which they produced a counterexample trace.

We do so by checking whether each counterexample trace is still causing a violation on the translated C implementation of the neural network. Throughout this process, we eliminate $2$ out of $208$ benchmarks with unsafe verdict. Specifically, we can not reproduce any of the five counterexamples for property $5$ of the Robot subcategory, and the six counterexamples for property $6$ of the VDP subcategory. As such, the \texttt{reach\_prob\_density} category in \bench{} contains only $35$ benchmarks out of the original $37$.

\section{Benchmark Format}
\label{sec:format}

In this section, we describe how to make a benchmark like \bench{2.0} compatible with most existing software verifiers. This is a crucial step of our methodology because verifiers are highly specialised tools and, as such, can be complex to configure and use correctly without understanding their internal implementation details. Fortunately, the international software verification competition (SV-COMP) \cite{beyer2024state} has already established a common format and verification workflow over the past decade. Our goal is to match it.\footnote{An early version of our benchmark, \bench{1.0}, is already part of the official benchmark set of SV-COMP (see Section \ref{sec:results_progress} for details).}

In SV-COMP, tools verify C programs against a specification. The specification can range from memory management to the absence of overflows or simply ensuring that a specific statement is not reachable.\footnote{The competition rules are available at \url{https://sv-comp.sosy-lab.org/2025/rules.php}.} The latter category is called \textit{ReachSafety} in SV-COMP parlance and includes a large set of verification instances whose goal is to check that a custom error function (i.e. \texttt{reach\_error}) is never invoked. A neural network verification benchmark like \bench{2.0}, with its explicit pre-condition and post-condition structure (see Equation \ref{eq:safety_prop}), falls into this category.

\begin{figure}[t]%
\centering
    \includegraphics[width=\textwidth]{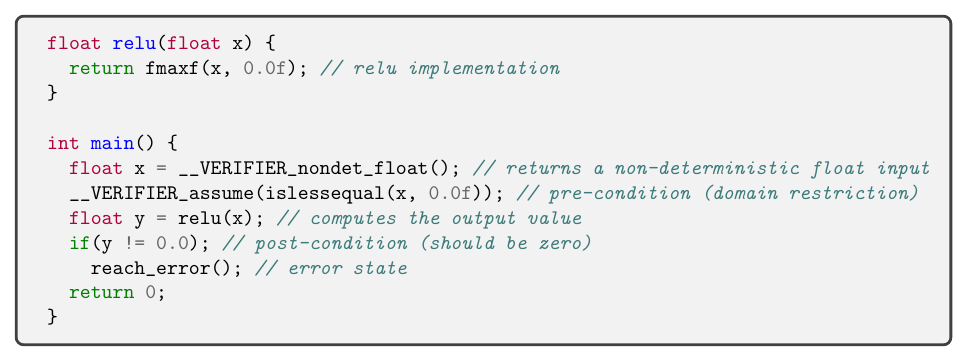}
\caption{Example of \textit{ReachSafety} harness. The property requires that applying ReLU over a non-positive input must return zero.}%
\label{fig:harness-example}%
\end{figure}

Figure~\ref{fig:harness-example} contains an example of an SV-COMP harness (a C program with a property) in the \textit{ReachSafety} format. Note that the harness includes two auxiliary intrinsic functions: \texttt{\_\_VERIFIER\_nondet\_float()} and \texttt{\_\_VERIFIER\_assume(\_Bool)}. The first constructs a non-deterministic float variable\footnote{In this context, nondeterministic means that it can assume any value representable by the float type.}, while the second adds a pre-condition on its value. We explain the role of such intrinsic functions in Section \ref{sec:intrinsics}.

The \textit{ReachSafety} format we show in Figure~\ref{fig:harness-example} has a few advantages over the plain C format from our introductory example of Figure~\ref{fig:sv-example}. First, it provides an explicit mechanism for specifying pre- and post-conditions via intrinsic functions, thereby removing semantic ambiguity. Second, it allows us to apply the same evaluation methodology used in SV-COMP, including a highly reproducible benchmarking workflow via the BenchExec framework~\cite{beyer2019reliable}. Third, it lets us import the verifier settings that the tool authors have already optimised for competition purposes, thus ensuring a fair comparison among them.

The rest of the section lists the requirements of the SV-COMP format (Section \ref{sec:format_requirements}), presents our workflow to convert the benchmark files (Section \ref{sec:workflow}), and covers additional requirements to make \bench{2.0} compatible with BenchExec (Section \ref{sec:bench-requirements}).

\subsection{SV-COMP Requirements}
\label{sec:format_requirements}

The verification benchmarks used in SV-COMP must satisfy the following requirements:
\begin{itemize}
    \item \textit{No Undefined Behaviour.} The benchmarks must not rely on any undefined behaviour of the C standard. For example, all variables must be initialized and read and write operations must not contain buffer overflows. Furthermore, all functions need to be declared and defined, unless they are part of the C standard itself. The only way to introduce undefined behaviour is via explicit intrinsic functions (see Section \ref{sec:intrinsics}).
    \item \textit{Self-Contained File.} In real-world C programs, it is common to split the implementation among multiple files and libraries. For SV-COMP purposes, the entire program must fit into a single source file. Additionally, only standard C headers are allowed for inclusion (e.g. \texttt{stdlib.h}, \texttt{math.h}).
    \item \textit{Explicit Host Configuration.} In the C standard, certain implementation details are left to the specific system architecture the program is compiled on. In contrast, a  SV-COMP benchmark should not rely on any implementation of the standard (X86 vs ARM, Windows vs Linux). If a specific behaviour is required, it needs to be explicitly defined in the benchmark.     
\end{itemize}
Among these requirements, the first requires manual inspection, the second can be automatically checked with a compiler, and the third usually involves running a compiler preprocessor in the host system to generate the source file with all the required characteristics. We present our approach to satisfy them in Section \ref{sec:workflow}.

\subsubsection{Intrinsic Functions}
\label{sec:intrinsics}

Some verification benchmarks are designed to reason on the consequences of undefined behaviour. For example, a benchmark might want to use an undefined variable to simulate an input from the user or a read from a sensor. In order to model these situations, SV-COMP introduces a number of intrinsic functions. In general, they can be thought of as extensions to \texttt{libc} that allow us to control the thread scheduling, specify pre-conditions, and set nondeterministic values, among others. Here, we describe the ones that are relevant to our work:
\begin{itemize}
    \item \textit{Nondeterministic Values.} The function \texttt{\_\_VERIFIER\_nondet\_X(void)} returns a nondeterministic value of type \texttt{X}, where \texttt{X} is any C primitive type, such as \texttt{int} or \texttt{float}. In this context, nondeterministic means any value that the type can be assigned to.
    \item \textit{Preconditions.} The function \texttt{\_\_VERIFIER\_assume(\_Bool)} is used to define a path precondition. It only allows the execution to continue if the condition holds. Otherwise, it causes the path of execution to be unreachable (and vacuously correct).
    \item \textit{Violation Functions.} For the \textit{ReachSafety} category, the intrinsic function \texttt{reach\_error(void)} specifies that the path contains an error. If a program can reach this statement, it implies that the property has been violated.
\end{itemize}
We give an example of the use of these intrinsic functions in Figure \ref{fig:harness-example}.

\subsection{Benchmark Conversion Workflow}
\label{sec:workflow}

\begin{figure}[t]
  \centering
  \includegraphics[width=0.9\linewidth]{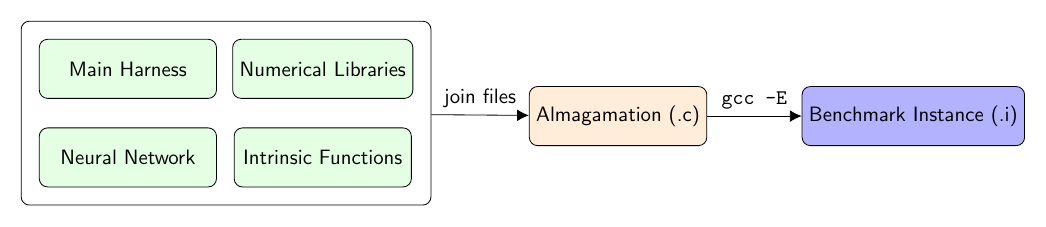}
  \caption{Benchmark conversion workflow.}
  \label{fig:benchmark-generation}%
\end{figure}

With all the requirements defined in Section \ref{sec:intrinsics}, we can describe our workflow to automatically convert our benchmarks to the SV-COMP format (see Figure \ref{fig:benchmark-generation}). To minimise duplication, the \bench{} repository\footnote{\url{https://github.com/emanino/plain_c_nn_benchmark}} divides the source code of each benchmark instance into the following components:
\begin{itemize}
    \item \textit{Main Harnesses.} These files are the entry point of the C program. All of them provide a \texttt{main()} function with pre- and post-conditions, as shown in Figure~\ref{fig:harness-example}. We make sure that all instances of undefined behaviour are specified via calls to the SV-COMP intrinsic functions (see Section \ref{sec:intrinsics}).
    \item \textit{Neural Networks.} \bench{2.0} categories that involve whole neural networks (see Table \ref{tab:nn_arch}) tend to reuse the same neural network code for multiple safety properties. For this reason, we keep the neural network implementation in a separate file.
    \item \textit{Numerical Libraries.} All benchmarks generated with the help of keras2c (see Table \ref{tab:nn_arch}) require an additional library that provide the implementation of Keras layers. Furthermore, experiments that involve the use of explicit operational models for \texttt{math.h} rely on linking with either the \musl{} or \coremath{} libraries (see Section \ref{sec:op_models}).
    \item \textit{Intrinsic Functions.} These files contain the headers with the intrinsic function declarations (see Section \ref{sec:intrinsics}).
\end{itemize}

\begin{figure}[t]%
\centering
    \includegraphics[width=\textwidth]{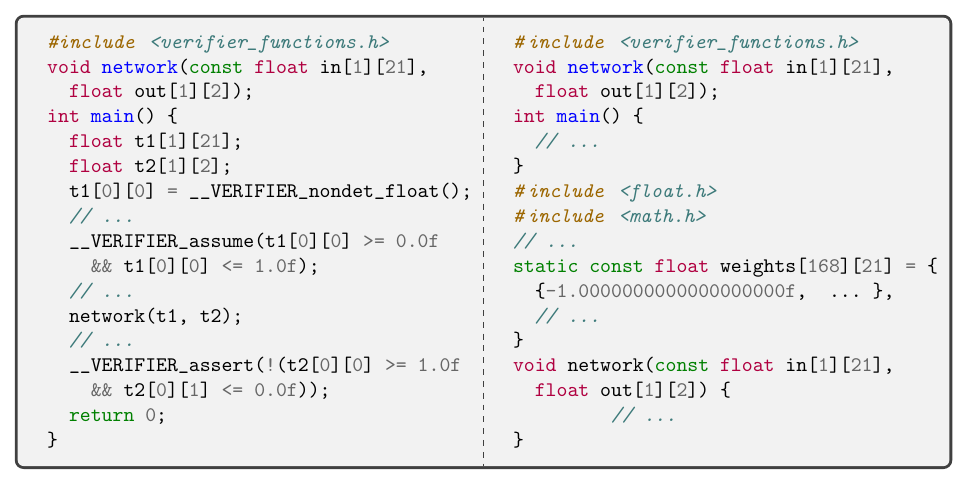}
\caption{Example of amalgamation from \texttt{prop\_bool\_v21\_c126\_safe.c}. The main harness (left) is combined with the neural network implementation, numerical libraries and intrinsic functions (right). For reason of space, the code shown here has been greatly simplified.}%
\label{fig:almagamation-example}%
\end{figure}

As a first step, we need to merge each harness with all its dependencies and produce a single self-contained file (see Section~\ref{sec:format_requirements}). We do so via a so-called \textit{amalgamation} stage. The term is inspired by the SQLite project~\cite{sqliteSQLiteAmalgamation}, where an amalgamation is a single C source file that contains everything needed to compile the library. For \bench{2.0}, we automate this process with CMake.\footnote{https://cmake.org/} We show an example of the result in Figure~\ref{fig:almagamation-example}.

As a second step, we need to make explicit the host configuration of the self-contained source file (see Section~\ref{sec:format_requirements}). On the one hand, we include all relevant library headers and preprocess the resulting file with the GCC compiler.\footnote{\url{https://gcc.gnu.org/onlinedocs/gcc/Function-Attributes.html}} On the other hand, we need to remove any system-specific or compiler-specific intrinsics. For the latter, we notice that GCC tends to introduce a number of function attributes to mark them as allocations. Since attributes are not standard across compilers, we use the Clang compiler\footnote{\url{https://clang.llvm.org/docs/AttributeReference.html}} to identify them before removal.

As a final sanity check step, we make sure that we are still able to compile the benchmarks with GCC. During this check, we use mock-up implementations of the SV-COMP intrinsic functions to avoid linking errors.

\subsection{BenchExec Setup}
\label{sec:bench-requirements}

In addition to the \texttt{.i} sources containing the C program and \textit{ReachSafety} specification, a well-formed SV-COMP benchmark provides additional files in four different formats. These files are consumed by the BenchExec framework~\cite{beyer2019reliable} when running the software verifiers, and allow us to automate the whole evaluation process. More specifically, the four file formats specify the following information:
\begin{itemize}
    \item \textit{Tool Definition.} An \texttt{.xml} file for each software verifier. The files specify the resource limits and options used by the respective tool. Additionally, they list which run-sets should be used for the analysis. We use the same files submitted by tool authors to the \textit{ReachSafety} category of SV-COMP and replace the run-sets with those of \bench{2.0}.
    \item \textit{Run-Set.} A \texttt{.set} file that lists all benchmark instances in a given category. The file contains a regex that matches all relevant source code files.
    \item \textit{Specification.} A \texttt{.yml} file for each benchmark instance. Each file lists the source code (see Section \ref{sec:workflow}), the word size ($32$ or $64$ bits), the property definition, and the expected verdict (safe or unsafe). We produce these files in a programmatic way.
    \item \textit{Property.} A \texttt{.prp} file with the definition of the safety property to be verified in LTL format. For this work, we use the same specification as the SV-COMP \textit{ReachSafety} category:
    \begin{equation}
    \label{eq:ltl_reach_safety}
        \mathtt{CHECK(\;init(main()),\;LTL(G\;!\;call(reach\_error()))\;)}
    \end{equation}
    which states that, when starting the program execution from the \texttt{main} function, the function \texttt{reach\_error} should never be invoked.
\end{itemize}

With this setup, we can run BenchExec on \bench{2.0} and evaluate all software verifiers that participate in SV-COMP.  We do so in Section \ref{sec:exp}.

\section{Experiments}
\label{sec:exp}

\textcolor{black}{After designing \bench{} (Section \ref{sec:bench}), and building it in a format compatible with existing automated software verifiers (Section \ref{sec:format}), we can now run our experimental study on the verification of floating-point neural networks at the software level.} To do so, we take the perspective of an \textit{expert user}, who knows how to install, run, and analyse the results of automated verification tools. As such, we assume that we can only apply minimal to no configuration changes to existing verification tools. This is in direct contrast to \textit{tool authors}, who have deep knowledge about the configuration options (e.g. command-line flags) and can potentially introduce sizable changes to improve the performance of their tool.\footnote{Note the all authors of this paper are involved with the development of ESBMC~\cite{Menezes2024esbmc}. However, we explicitly avoided making changes to ESBMC for the purposes of the present paper, treating it as an external tool instead (see Section \ref{sec:setup_tools}).}

In this light, we want to answer the following experimental questions using NeuroCodeBench:
\begin{itemize}
    \item \textit{Tool Correctness.} Can we trust the verdicts produced by automated software verifiers? Thanks to the presence of ground-truth labels for each \bench{} instance, we can check whether verifiers return the correct verdict. That gives us an indication of whether these tools can be used as a safety oracle for the floating-point implementations of neural networks. The results are given in Section \ref{sec:results_plain}.
    \item \textit{Maths Library Support.} Does the presence of calls to \texttt{math.h} affect automated software verifiers? A large proportion of \bench{} instances contains functions calls to \texttt{math.h}, which the verifiers might not support. This gives us an opportunity to test whether it is possible to provide external definitions (operational models) of these functions to the verifiers. The results are given in Section \ref{sec:results_maths}.
    \item \textit{Scalability.} How large of a problem can existing automated verification tools solve? By design, \bench{} contains instances in a wide range of complexity, from individual layers and activations to full neural networks. With it, we can estimate the scalability of verification tools, which informs the potential applications for which they can be employed. The results are given in Section \ref{sec:results_complexity}.
    \item \textit{Historical Progress.} Has the performance of automated software verifiers improved in the past few years? Since \bench{} is a new benchmark, it can be used to retroactively investigate the evolution of verification tools and establish whether general improvements in software verification readily translate to floating-point neural network code. We do so by going as far back as 2018. The results are given in Section \ref{sec:results_progress}.
\end{itemize}

\textcolor{black}{Before covering the results of the reproducibility study, we discuss the details of our experimental setup in Section \ref{sec:setup}.}

\subsection{Experimental Setup}
\label{sec:setup}

Our experimental setup is designed to replicate the conditions of SV-COMP as much as possible, as they constitute one of the highest standards for scientific reproducibility in the community~\cite{beyer2019reliable,beyer2024state}, as detailed in Section \ref{sec:format}. For reference, we provide a public stand-alone copy of \bench{2.0} and the results of the verifier runs.\footnote{\url{https://zenodo.org/records/21873894}}

\subsubsection{Software Verifiers.}
\label{sec:setup_tools}

For our experiments, we compare eight state-of-the-art software verifiers: \twols{}~\cite{malik2018}, \cbmc{}~\cite{Kroening2014}, \cpachecker{}~\cite{baier2024}, \divine{}~\cite{baranov2017}, \esbmc{}~\cite{Menezes2024esbmc}, \pesco{}~\cite{richter2019}, \pinaka{}~\cite{chaudhary2019}, \uautomizer{}~\cite{heinzmann2013}. These eight tools have been taking part in SV-COMP for several years, are actively maintained by their respective research groups, and support C programs with floating-point computation. Other software verification tools exist~\cite{beyer2024state}, but we exclude them from our experiments due to a range of factors outside of our control spanning licensing restrictions, limited support for floating-point C programs, and local configuration issues. For reproducibility reasons, we use the versions available on the SV-COMP website\footnote{\url{https://sv-comp.sosy-lab.org/}} without modifications, and run them on \bench using a local instance of BenchExec $3.23$~\cite{beyer2019reliable}. When not otherwise specified, we use the SV-COMP'24 version of the tools. In the historical comparison of Section \ref{sec:results_progress}, we refer to the archived versions of ESBMC on the SV-COMP website from 2018 to 2025 (current year).

\subsubsection{Hardware Platform.}
\label{sec:setup_hardware}

We run all experiments on an Ubuntu 20.04 machine with Kernel 5.4.0-177-generic. The hardware is a 64-bit 32-core Intel(R) Xeon(R) CPU E5-2620 v4 @ 2.10GHz with 170GiB of RAM. We allow the verifiers to use 15GB of RAM, 900s of CPU time, and four cores per each benchmark instance. The bulk of our experiments took place in three main phases that spanned from July 2024 to June 2025, \textcolor{black}{followed by a few minor updates in June 2026.} In total, we estimate that reproducing the experiments in this paper would take approximately $3163$ CPU hours: $2312$ CPU hours for the results in Sections \ref{sec:results_plain}, \ref{sec:results_maths}, \ref{sec:results_complexity}, and $851$ CPU hours for the historical results of ESBMC (excluding 2024, since it is already included in the previous count) in Section \ref{sec:results_progress}.

\subsubsection{Implementation of \texttt{math.h}}
\label{sec:setup_maths}

We bundle with \bench{2.0} the plain C implementations of \texttt{math.h} provided by the \musl{} and \coremath{} libraries (see Sections \ref{sec:op_models} and \ref{sec:workflow}). Specifically, we extract all relevant 32-bit floating-point functions listed in Section \ref{sec:bench_maths}.\footnote{See \url{https://github.com/emanino/plain_c_nn_benchmark/tree/main/extern/math_op_models}} In the case of \coremath{}, we omit the \texttt{sqrtf} function, which is not provided, and the \texttt{sinf} and \texttt{cosf} functions. The latter two functions rely on 128-bit float arithmetic, which is not supported by the majority of existing software verifiers. Furthermore, we reduce the external dependencies of the \musl{} library to the minimum necessary for successful compilation, and replace the bit manipulation primitives in \coremath{} with plain C implementations. We also address minor issues with type casting and macro definitions.

\subsection{Tool Correctness}
\label{sec:results_plain}

Table \ref{tab:plain-verifiers-all-verdicts} provides a high-level overview of the output of the verifiers. After analysing each instance in \bench{2.0} for $900$ seconds, the verifiers return \textit{inconclusive} verdicts for the majority of the benchmark. More worryingly, a non-insignificant percentage of the verdicts are incorrect ($3\%$ in total).

\begin{table}[ht]
\caption{Breakdown of all verdicts returned by the verifiers for \bench{2.0} benchmarks ($912$ overall).}
\label{tab:plain-verifiers-all-verdicts}
\centering
\begin{tabular}{c|cc|cc|cccc}
    \toprule
    \multirow{2}{*}{Verifier} & 
    \multicolumn{2}{c|}{Correct} & 
    \multicolumn{2}{c|}{Incorrect} & 
    \multicolumn{4}{c}{Inconclusive}\\
    \cmidrule{2-9}
    & Safe & Unsafe & Safe & Unsafe & Timeout & Out of Memory & Unknown & Error \\
    \midrule
    \twols{}          & 12 & 2 & - & - & 4 & - & 894 & -\\
    \cbmc{}           & 66 & 305 & - & 121 & - & 80 & - & 340\\
    \cpachecker{}     & 18 & 19 & - & - & 292 & 265 & - & 318\\
    \divine{}         & 6 & 3 & - & - & 390 & 152 & 361 & -\\
    \esbmc{}          & 76 & 86 & - & - & 539 & 211 & - & -\\
    \pesco{}          & 24 & 25 & 10 & - & 482 & 58 & - & 313\\
    \pinaka{}         & 44 & 71 & - & 82 & 243 & 292 & - & 180\\
    \uautomizer{}     & 9 & 27 & - & 12 & 754 & 1 & 106 & 3\\
    \bottomrule
\end{tabular}
\end{table}

The only tool that produces the most \textit{definitive} verdicts (i.e., safe or unsafe) is \cbmc{}, with $492$ definitive against $440$ inconclusive verdicts. However, nearly a quarter of them are incorrect. Similarly, more than a third of all definitive verdicts produced by \pinaka{} and exactly a quarter of all definitive verdicts reported by \uautomizer{} are incorrect. In all three cases, these are incorrect \textit{unsafe} verdicts, i.e. the tools report the presence of a spurious bug. In contrast, the only tool that returns around $17\%$ of incorrect \textit{safe} verdicts is \pesco{}.

Although the other verifiers appear to produce only correct verdicts, we cannot exclude the possibility that some of them might be ``lucky guesses''. Indeed, given the balanced number of safe and unsafe instances in \bench{2.0} (see Table \ref{tab:bench_overview}), a hypothetical verifier that produces random verdicts would be correct $50\%$ of the time on average. The SV-COMP rules~\cite{beyer2024state} discourage guessing by penalising verdicts that are not supported by a valid proof. Unfortunately, this process of \textit{witness validation} relies on running an ensemble of specialised software verifiers to check the proof, which could still return a wrong result in our setting. For this reason, we choose to only evaluate the verifiers against our ground-truth verdicts.

With that caveat in mind, \cbmc{}, \esbmc{}, and \pinaka{} are the three tools that correctly solve the largest number of instances: $371$, $162$, and $115$ respectively. Among them, \esbmc{} is the only one that does not return incorrect verdicts. For this reason, we focus on \esbmc{} in our later analysis of scalability and historical progress (see Sections \ref{sec:results_complexity} and \ref{sec:results_progress}).

On a different note, all inconclusive verdicts in Table \ref{tab:plain-verifiers-all-verdicts} can be broadly placed into one of two subcategories: exhaustion of computational resources or defence mechanisms against unsupported features. For the former, the tools typically report a timeout or out of memory message. Occasionally, some of them return an error code instead. For example, \cbmc{} produces a specific error code instead of a timeout. For the latter, many tools resort to outputting an unknown or error verdict when they encounter a mathematical function that is not currently supported. We give a detailed analysis of log outputs and time distributions in Appendices  \ref{sec:observations} and \ref{sec:time_distro}.

\subsection{Maths Library Support}
\label{sec:results_maths}

One of the difficulties of verifying software that features calls to external libraries comes from the fact that a verifier will struggle to reason on them if the implementations of these libraries are not provided. Although most verifiers support the core features of the C language, including many standard libraries, the mathematical functions in \texttt{math.h} are often overlooked. Here, we measure how verifiers perform on benchmark instances that contain calls to these functions and we evaluate whether introducing explicit C implementations of \texttt{math.h} improves the verification outcomes.

\begin{figure}[t]
\centering
    \begin{subfigure}[b]{0.49\textwidth}
        \includegraphics[width=\textwidth]{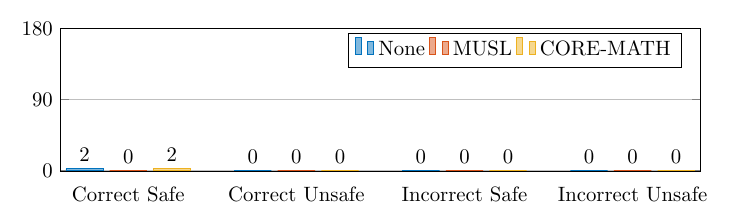}
        \caption{\twols{}}
        \label{fig:twols_math}
    \end{subfigure}
    \hfill
    \begin{subfigure}[b]{0.49\textwidth}
        \includegraphics[width=\textwidth]{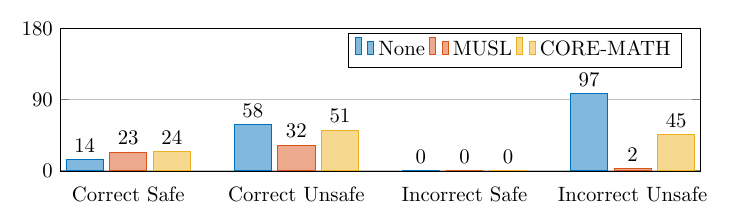}
        \caption{\cbmc{}}
        \label{fig:cbmc_math}
    \end{subfigure}
    \hfill\begin{subfigure}[b]{0.49\textwidth}
        \includegraphics[width=\textwidth]{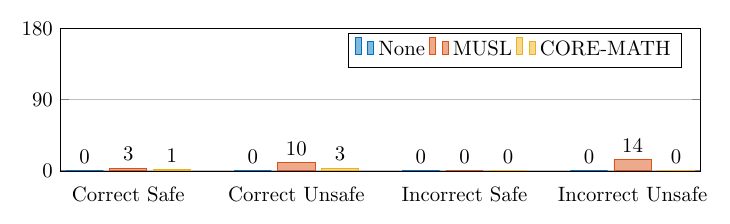}
        \caption{\cpachecker{}}
        \label{fig:cpa_math}
    \end{subfigure}
    \hfill\begin{subfigure}[b]{0.49\textwidth}
        \includegraphics[width=\textwidth]{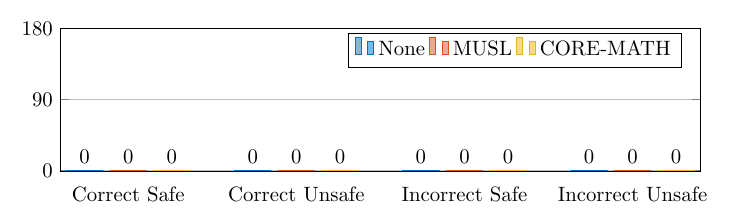}
        \caption{\divine{}}
        \label{fig:divine_math}
    \end{subfigure}
    \hfill\begin{subfigure}[b]{0.49\textwidth}
        \includegraphics[width=\textwidth]{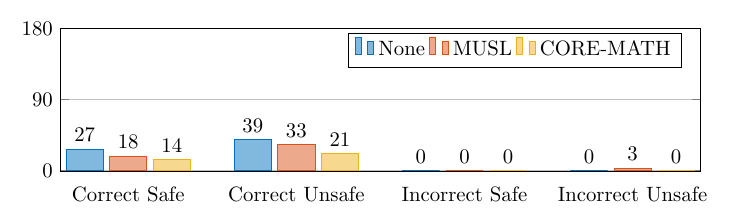}
        \caption{\esbmc{}}
        \label{fig:esbmc_math}
    \end{subfigure}
    \hfill\begin{subfigure}[b]{0.49\textwidth}
        \includegraphics[width=\textwidth]{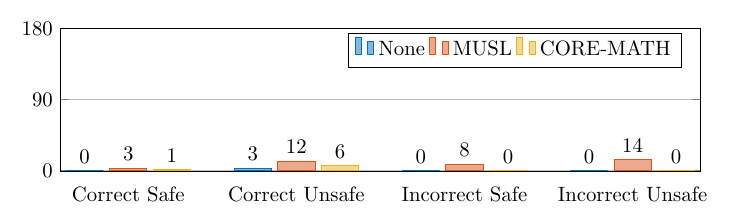}
        \caption{\pesco{}}
        \label{fig:pesco_math}
    \end{subfigure}
    \hfill\begin{subfigure}[b]{0.49\textwidth}
        \includegraphics[width=\textwidth]{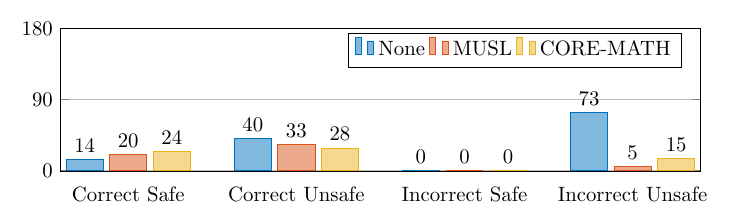}
        \caption{\pinaka{}}
        \label{fig:pinaka_math}
    \end{subfigure}
    \hfill\begin{subfigure}[b]{0.49\textwidth}
        \includegraphics[width=\textwidth]{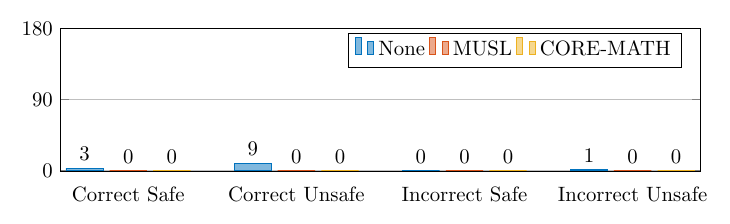}
        \caption{\uautomizer{}}
        \label{fig:uauto_math}
    \end{subfigure}
\caption{Verifier performance on the $180$ benchmark instances of \bench{2.0} that contain \texttt{math.h} functions.}
\label{fig:math-funs}
\end{figure}

To do so, we isolate the $180$ instances that contain calls to \texttt{math.h} by filtering \bench{2.0} against the list of mathematical functions in Table \ref{tab:math-h-functions}. Then, we concatenate the \musl{} or \coremath{} operational models to each benchmark instance as detailed in Section \ref{sec:workflow}. Table \ref{tab:math-funs} and Figure \ref{fig:math-funs} compare the performance of software verifiers with and without \texttt{math.h} operational models.

\begin{table}[t]
\caption{Definite verdicts (both correct and incorrect) on the subset of \bench{2.0} with calls to \texttt{math.h} ($180$ instances).}
\label{tab:math-funs}
\centering
\begin{tabular}{l|cccccccc}
     \toprule
     Op. Models &  
     \rotatebox[origin=c]{90}{\twols{}} &
     \rotatebox[origin=c]{90}{\cbmc{}} &
     \rotatebox[origin=c]{90}{\cpachecker{}} &
     \rotatebox[origin=c]{90}{\divine{}} &
     \rotatebox[origin=c]{90}{\esbmc{}} &
     \rotatebox[origin=c]{90}{\pesco{}} &
     \rotatebox[origin=c]{90}{\pinaka{}} &
     \rotatebox[origin=c]{90}{\uautomizer{}}\\
     \midrule
     None       & 2 & 169 &  - & - & 66 &  3 & 127 & 13 \\
     \musl{}     & - &  57 & 27 & - & 54 & 37 &  58 &  - \\
     \coremath{} & 2 & 120 &  4 & - & 35 &  7 &  67 &  - \\
     \bottomrule
\end{tabular}
\end{table}

Overall, the introduction of operational models has opposite effects on different software verifiers, as Table \ref{tab:math-funs} shows. On the one hand, verifiers like \cbmc{}, \esbmc{}, \pinaka{}, and \uautomizer{} see a reduction in the number of solved instances. We believe this is a consequence of the additional complexity of reasoning on the implementation of mathematical functions, i.e., there is more C code to verify when the libraries are provided. On the other hand, tools like \cpachecker{} and \pesco{} greatly benefit from the presence of operational models. In their case, we believe that \musl{} yields better results than \coremath{} because the latter implementations often rely on double-precision arithmetic followed by truncation, instead of performing the computation in single-precision directly.

At the same time, our goal is not just to increase the total number of solved instances, but rather the number of \textit{correct} ones. In this respect, Figure \ref{fig:math-funs} shows a more detailed picture. Specifically, the majority of verdicts produced by \cpachecker{} and \pesco{} with the \musl{} operational models turns out to be incorrect. For \pesco{}, the increase includes incorrect unsafe verdicts, which the tool never produces on the plain benchmark (see Table \ref{tab:plain-verifiers-all-verdicts}).

In contrast, the introduction of operational models benefits \cbmc{} and \pinaka{} the most. Indeed, these two tools see a significant reduction in the number of incorrect verdicts, with almost no impact on the total number of correct ones. In fact, \musl{} demonstrates the most striking effect by almost completely eliminating all incorrect verdicts.  Regarding \esbmc{}, the introduction of operational models resulted in a moderate reduction in performance.

For completeness, we also report the results on the $732$ benchmark instances that do \textit{not} contain relevant calls to \texttt{math.h} in Figure \ref{fig:no-math-funs}, together with the full result matrix in Table \ref{tab:verifiers-all-verdicts}. As that figure shows, the verifier performance is largely unaffected by the addition of (unused) operational models. This is not surprising, since most software verifiers employ slicing techniques to ignore the portions of code that do not have any effect on the verification conditions.

\subsection{Scalability}
\label{sec:results_complexity}

One of our design requirements for \bench{2.0} was to provide benchmark instances in a wide range of complexity (see Section \ref{sec:design}). With them, we can measure the scalability of software verifiers as the reasoning tasks become computationally harder. Table \ref{tab:subcategories} compares the performance of the verifiers between different subcategories, without operational models of \texttt{math.h} functions. There, we highlight the number of both correct and incorrect definitive verdicts (safe or unsafe) returned by each verifier. Also, we score the complexity of each category by computing a solution rate, which represents the average proportion of benchmarks solved by the verifiers.

\begin{table}[ht!]
\caption{Number of correct (incorrect) verdicts returned by the verifiers for each category of \bench{2.0}. We sort the categories by their average solution rate, i.e. the percentage of instances solved correctly (incorrectly) by the average verifier.}
\label{tab:subcategories}
\centering
\begin{tabular}{l|cccccccc|c}
     \toprule
     Benchmark Category &  
     \rotatebox[origin=c]{90}{\twols{}} &
     \rotatebox[origin=c]{90}{\cbmc{}} &
     \rotatebox[origin=c]{90}{\cpachecker{}} &
     \rotatebox[origin=c]{90}{\divine{}} &
     \rotatebox[origin=c]{90}{\esbmc{}} &
     \rotatebox[origin=c]{90}{\pesco{}} &
     \rotatebox[origin=c]{90}{\pinaka{}} &
     \rotatebox[origin=c]{90}{\uautomizer{}} &
     \rotatebox[origin=c]{90}{Solution Rate}\\
     \midrule
     Neural Layers ($n = 86$) & - & 69 (7) & 27 & 5 & 72 & 36 & 59 (13) & 23 (8) & 42\% (4\%)\\
     Activation Functions ($n = 57$) & 12 & 23 (33) & 10 & 4 & 27 & 11 & 19 (36) & 8 (4) & 25\% (16\%) \\
     Maths Functions ($n = 58$) & 2 & 22 (33) & - & - & 25 & - & 22 (33) & 5 & 16\% (14\%)\\
     \midrule
     Lipschitz-Bounded Nets ($n = 108$) & - & 55 & - & - & 16 & - & 14 & - & 9.8\% \\
     SAT ReLU Networks ($n = 96$) & - & 41 (24) & - & - & 16 & (10) & 1 & - & 7.6\% (4.4\%)\\
     Polynomial Approx. ($n = 96$) & - & 45 & - & - & 6 & - & - & - & 6.6\% \\
     Hopfield Networks ($n = 80$) & - & 36 (24) & - & - & - & 2 & - & - & 5.9\% (3.8\%)\\
     \midrule
     Reinforcement Learning ($n = 296$) & - & 80 & - & - & - & - & - & - & 3\% \\
     Probability Density ($n = 35$) & - & - & - & - & - & - & - & - & 0\% \\
     \bottomrule
\end{tabular}
\end{table}

As expected, the three easiest categories to solve are those that contain individual components of a neural network: neural layers, activation functions, and mathematical functions. Among them, neural layers involves the lowest number of calls to \texttt{math.h}, with only $44\%$ of instances, compared to $77\%$ and $100\%$ for the other two categories, respectively. As a consequence, the number of incorrect verdicts in neural layers remains relatively small.

As we move onto benchmark instances that contain whole neural networks, the solution rate drops into the single digits, and many state-of-the-art software verifiers cease to return useful output. Still, we can see a difference between synthetically-generated benchmarks and real-world ones. Indeed, we constructed the former four categories (i.e., Lipschitz-Bounded Networks, SAT ReLU Networks, Polynomial Approximation, Hopfield Networks) to include a few small instances (see Section \ref{sec:bench}), which the verifiers can solve successfully. In contrast, we converted the latter two categories (i.e. Reinforcement Learning and Probability Density) from the VNN-COMP competition format (see Section \ref{sec:bench_vnn_comp}), and their relatively larger neural network size ($64$ to $512$ neurons) is at the upper limit of what state-of-the-art software verifiers can handle.

In the remainder of the section, we identify the exact scalability limit of software verifiers by looking at the individual results of each benchmark instance. More specifically, we focus on the three subcategories containing synthetic networks (i.e., Lipschitz-Bounded Networks, SAT ReLU Networks, Polynomial Approximation Networks), and we report how much CPU time and memory the verifiers use. We mainly base our analysis on \esbmc{}, as it is the tool with the highest number of correctly solved benchmarks that produces no incorrect verdicts.

\subsubsection{Lipschitz-Bounded Networks}
\label{sec:results_complexity_lipschitz}

As we explain in Section \ref{sec:bench_lipschitz}, two complexity factors are at play in this subcategory. First, the network architecture itself varies in the number of inputs $i$ and the width of the hidden layers $w$, while the number of outputs $o=1$ and hidden layers $d=2$ remains the same. Due to the specific Lipschitz-bounded architecture we chose, the number of network parameters is $(2d-1)w^2+(i+o+d)w+o$, which scales quadratically in the network width $w$. Second, each network is associated with six different safety properties (cases). These differ in their decidability threshold: three are safe and three are unsafe. Cases $2,3$ are the closest to the threshold and thus the most difficult to solve, at least in theory.

 \begin{figure}[t]
    \includegraphics[width=\linewidth]{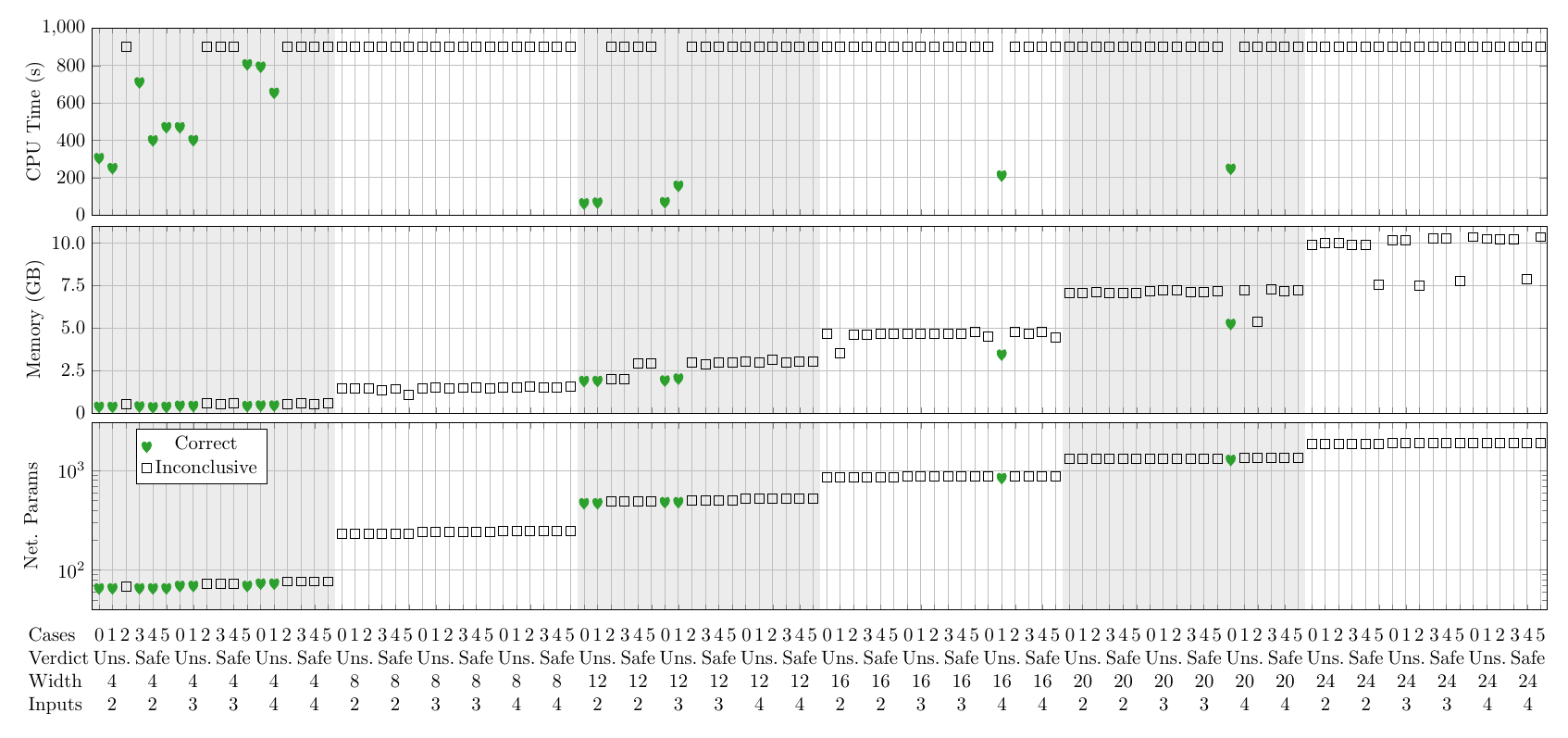}
    \caption{Performance of \esbmc{} on the Lipschitz-Bounded Networks category.}
    \label{fig:lipschitz-esbmc}
\end{figure}

Figure \ref{fig:lipschitz-esbmc} shows that \esbmc{} solves $16$ out of $108$ benchmark instances. Most of them ($10$ out of $16$) are the smallest networks with layer width $w=4$. As the layer width increases, \esbmc{} begins experiencing a higher rate of timeouts. Interestingly, all solved instances beyond $w>4$ are unsafe, i.e. \esbmc{} needs only to find one counterexample. Moreover, there is only one instance of cases $2,3$ that \esbmc{} is able to solve, thus confirming that being closer to the decision threshold makes for more challenging instances regardless of network size. In terms of memory consumption, \esbmc{} demonstrates a steady increase in memory consumption as the size of each layer in the networks increases. 

\subsubsection{SAT ReLU Networks}
\label{sec:results_complexity_sat_relu}

As we explain in Section \ref{sec:bench_gadgets}, the size of the networks in this category is directly correlated with the number of input variables $v$ and clauses $c$ in the original $CNF$ formula. More specifically, the number of parameters in the network is given by $(2v+c)(v+3)+2$, which scales quadratically in the number of input variables. We also expect unsafe instances to be easier to solve than safe ones, as finding a counterexample suffices.

\begin{figure}[t]
    \includegraphics[width=\linewidth]{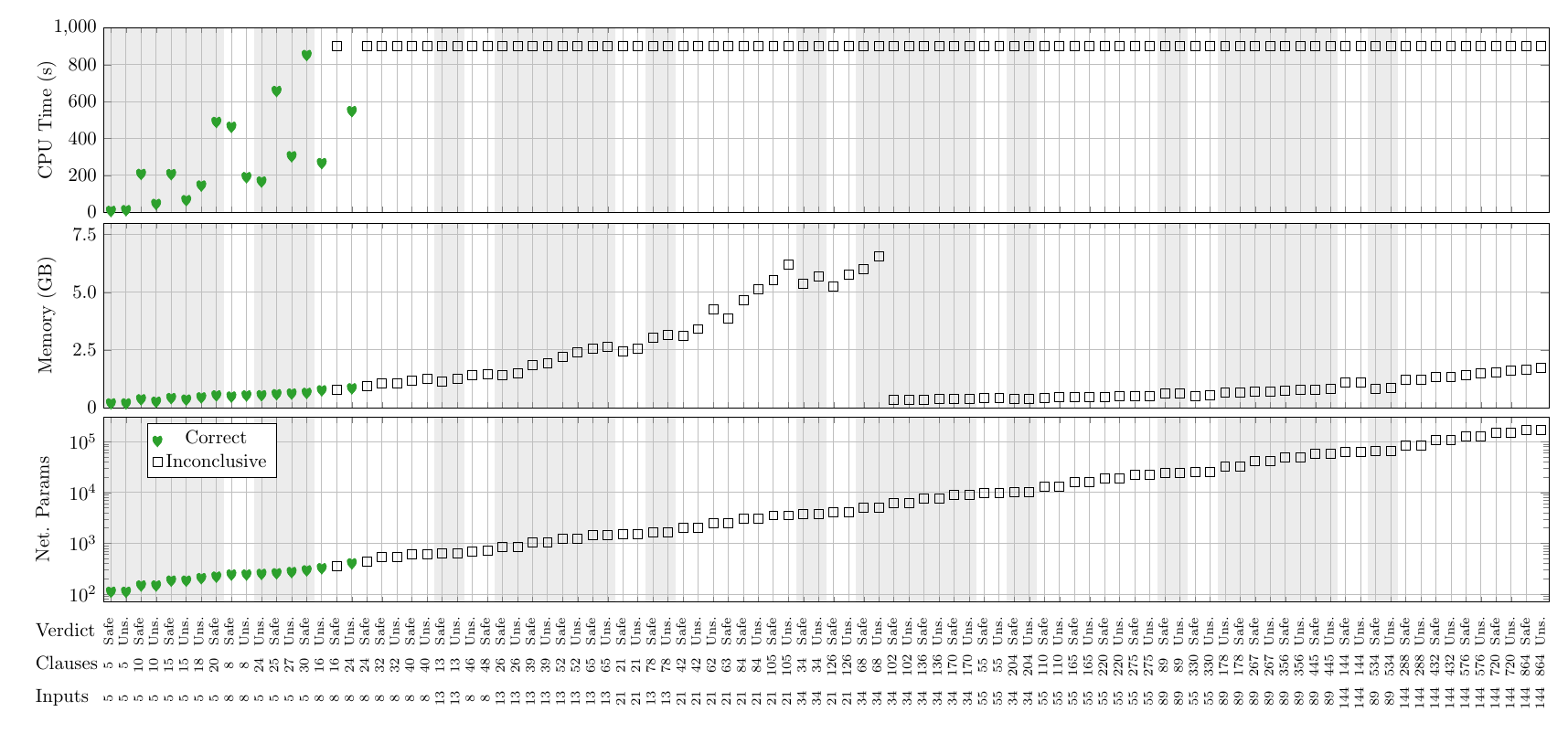}
    \caption{Performance of \esbmc{} on the SAT ReLU Networks category.}
    \label{fig:sat-relu-esbmc}
\end{figure}

Figure \ref{fig:lipschitz-esbmc} shows that \esbmc{} solves $16$ out of $96$ instances. They include all networks with only $v=5$ variables and four networks with $v=8$ variables. As expected, safe instances take considerably longer to solve and are the first to time out as the network size grows larger. As per the memory, we begin to observe a clear phenomenon: while the memory consumption initially increases with the complexity of the instances, it drops to zero again for instances with more than $6$K network parameters. We believe this phenomenon can be attributed to the bounded model checking nature of \esbmc{}, where the verification process consists of iterating between symbolic execution and SMT solving. Thus, reaching the timeout at different stages will result in different memory consumption profiles (e.g., symbolic execution is typically less memory intensive than SMT solving).

\subsubsection{Polynomial Approximation Networks}
\label{sec:results_complexity_poly_approx}

As we explain in Section \ref{sec:bench_poly}, the complexity of the present category is driven by two main factors. First, the number of network parameters can be computed as $(d-1)*w^2+(d+2)w+1$, where $d$ is the number of hidden layers and $w$ is their width. As such, the network size scales quadratically in the layer width $w$. Second, there are six safety properties (cases) for each network, three safe and three unsafe. The middle cases (i.e. $2,3$) are the closest to the decision threshold. Similar to Section \ref{sec:results_complexity_lipschitz}, we expect them to be more difficult to solve than the others.

\begin{figure}[t]
    \centering
    \includegraphics[width=\linewidth]{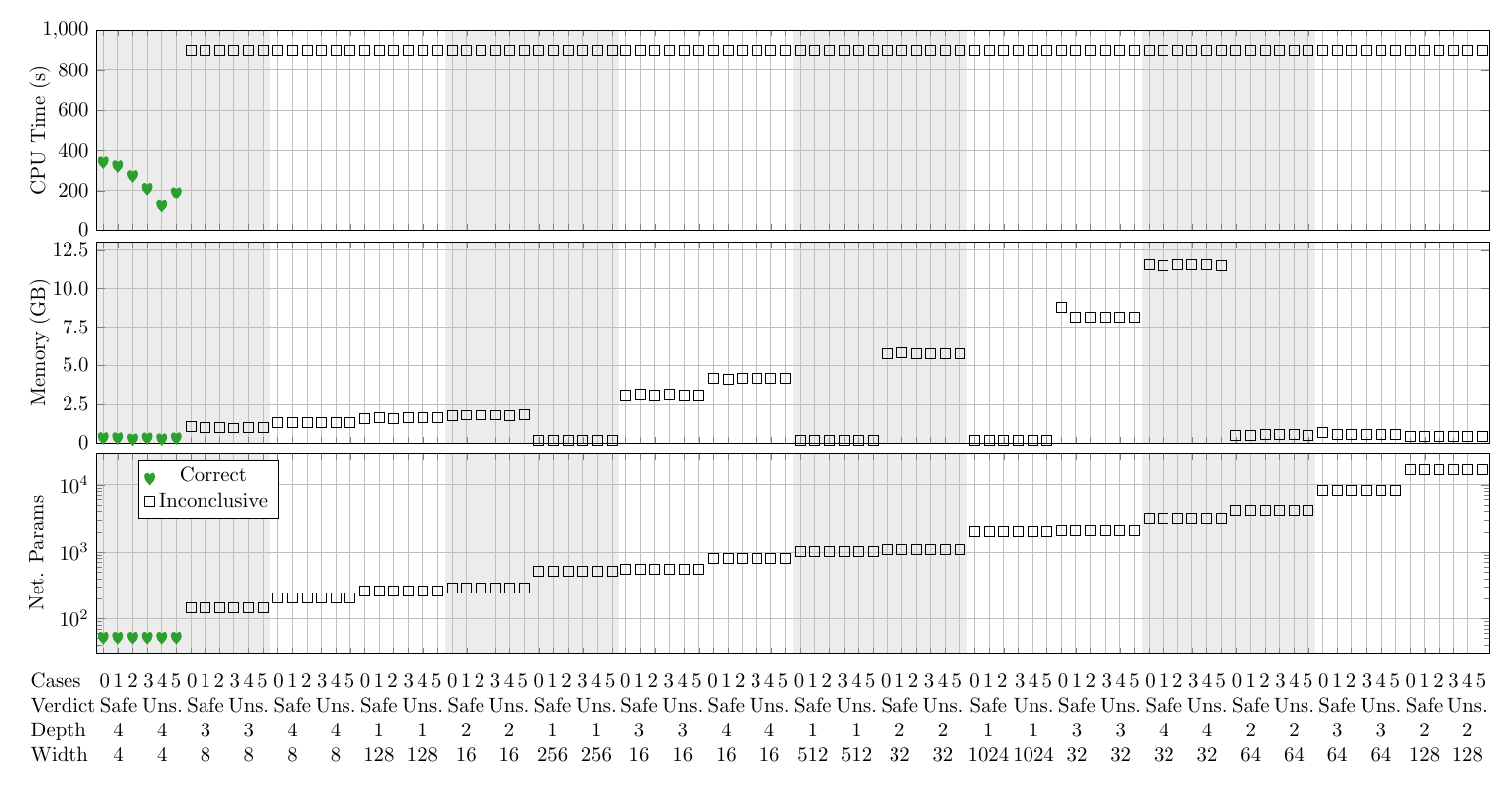}
    \caption{Performance of \esbmc{} on the Polynomial Approximation category.}
    \label{fig:poly-esbmc}    
\end{figure}

Figure \ref{fig:poly-esbmc} shows that \esbmc{} solves only the $6$ simplest benchmark instances, with the rest reaching timeout. Among these, we do not observe the expected peak of complexity around cases $2,3$. Instead, all unsafe instances appear easier to solve. In terms of memory consumption, \esbmc{} exhibits the same sudden drops as the SAT ReLU category (see Section \ref{sec:results_complexity_sat_relu}).

\subsection{Historical Progress}
\label{sec:results_progress}

One of the main inspirations behind our work were anecdotal observations of incorrect verdicts from software verifiers. Early development of \bench{} allowed us to confirm our observations, measure their frequency, and begin the process of engaging with the software verification community. This section examines the recent progress in the field.

\subsubsection{Impact on Tool Development}
\label{sec:results_progress_early}

We first introduced \bench{1.0}, an earlier version of our benchmark, into the SV-COMP benchmark suite in time for its 2024 edition~\cite{beyer2024state}. This was a crucial step in attracting the attention of the community to the problem of verifying C programs with extensive use of floating-point arithmetic and mathematical functions. Given the rules of SV-COMP, tool authors had a chance to look at the benchmark and patch their software verifiers before the competition. 

\begin{figure}[htb]
\centering
    \begin{subfigure}[b]{0.49\textwidth}
        \includegraphics[width=\textwidth]{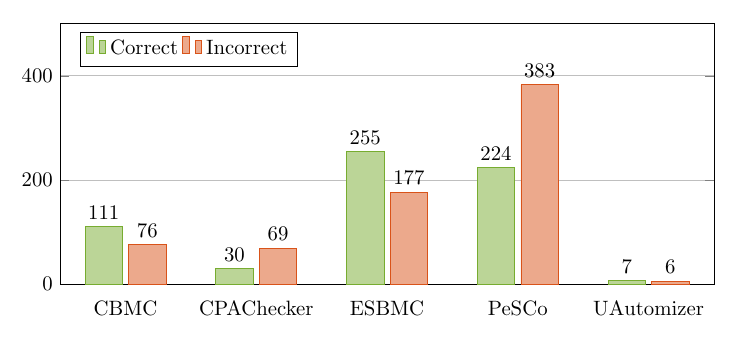}
        \caption{August 2023.}
        \label{fig:history_before}
    \end{subfigure}
    \hfill
    \begin{subfigure}[b]{0.49\textwidth}
        \includegraphics[width=\textwidth]{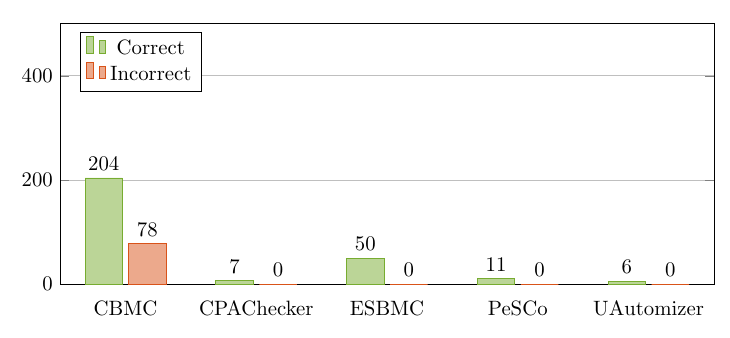}
        \caption{December 2023.}
        \label{fig:history_after}
    \end{subfigure}
\caption{Verifier performance on \bench{1.0} before and after its introduction into the SV-COMP benchmark suite.}
\label{fig:history}
\end{figure}

\begin{figure}[htb]
\centering
    \begin{subfigure}[b]{0.49\textwidth}
        \includegraphics[width=\textwidth]{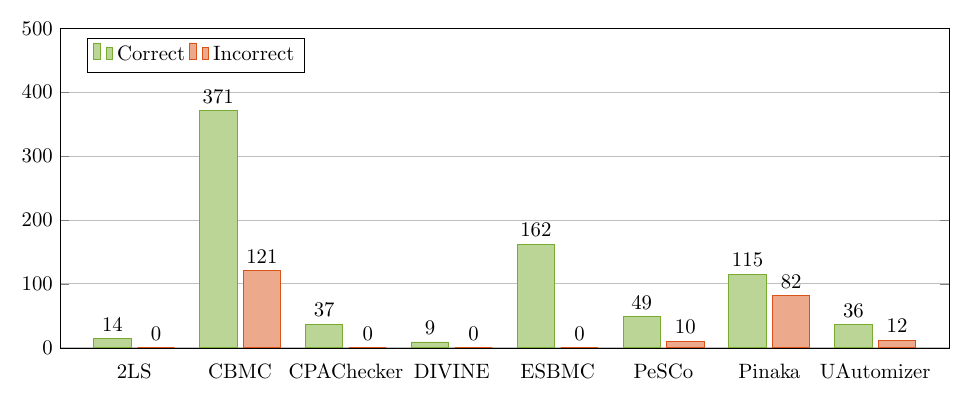}
        \caption{June 2025.}
        \label{fig:history_2_before}
    \end{subfigure}
    \hfill
    \begin{subfigure}[b]{0.49\textwidth}
        \includegraphics[width=\textwidth]{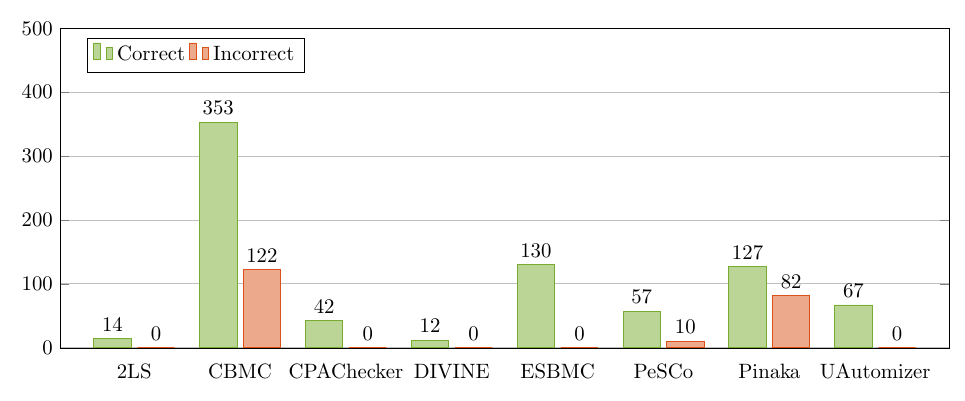}
        \caption{January 2026.}
        \label{fig:history_2_after}
    \end{subfigure}
\caption{Verifier performance on \bench{2.0} before and after its introduction into the SV-COMP benchmark suite.}
\label{fig:history_2}
\end{figure}

For this reason, the difference in tool performance before and after the introduction of \bench{1.0} is stark, as shown in Figure \ref{fig:history}. Note that the results in Figure \ref{fig:history_before} are our own, as detailed our preliminary technical report~\cite{manino2023neurocodebench}. In contrast, the results in Figure \ref{fig:history_after} come from the official SV-COMP 2024 runs, which we filtered to show the \bench{1.0} instances only. Informal communication with tool authors confirmed that a large portion of incorrect results was fixed in the run up to the competition in direct response to the release of our benchmark.

The newest version of our benchmark introduces significantly more instances and fixes some known issues, as detailed in Appendix \ref{sec:version_diff}. As such, it may reveal further incorrect behaviour in existing verifiers and drive further development of these tools. 

\textcolor{black}{It is important to note that all of the experiments in this paper were conducted before \bench{2.0} was included into the most recent addition of SV-COMP 2026. Since then some verifiers demonstrated some improvements (see Figure \ref{fig:history_2_after}). Although, the overall effect was not as noticeable as when \bench{1.0} was first introduced into SV-COMP 2024.}

\subsubsection{Performance Trends since 2018}
\label{sec:results_progress_history}

\bench{2.0} can also be used as a measuring tool to gauge the progress of software verifiers on floating-point reasoning over the years. In this section, we recover the executables and setups of older software verification tools from the historical records of SV-COMP. We cover the \textcolor{black}{nine} years spanning between the 2018 and \textcolor{black}{2026} editions of the competition. For simplicity, we study the performance of \esbmc{} only. We choose this specific tool because it has the best performance among those that do not produce any incorrect verdicts (see Table \ref{tab:plain-verifiers-all-verdicts}).

\begin{figure}[t]
    \centering
    \includegraphics[width=0.75\linewidth]{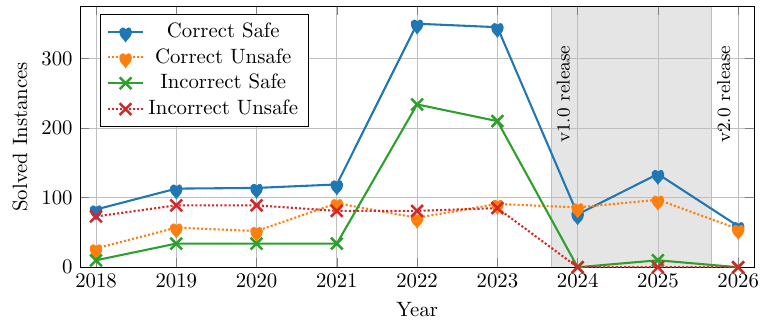}
    \caption{\bench{2.0} performance of different versions of \esbmc{} between 2018 and 2026.}
    \label{fig:esbmc_history}
\end{figure}

The results in Figure \ref{fig:esbmc_history} show a sharp increase in the number of safe verdicts around 2022 and 2023, including a large portion of incorrect ones. An internal investigation within our research group revealed that the increase was due to an erroneous implementation of an unrelated verification algorithm in \esbmc{} (i.e., $k$-induction), which remained undiscovered until the release of \bench{1.0}. The latter also caused a significant shift in development focus towards ensuring the verifier correctness on floating-point benchmarks in 2024. Indeed, after the release of \bench{1.0} the number of incorrect verdicts drops from $295$ to zero.

Additionally, there is a gradual improvement in the number of instances correctly solved by \esbmc{}, if we ignore the 2022-2023 spike. The trend continues in 2025, when \esbmc{} has yet again increased the number of correct instances from $162$ to $231$, albeit at the price of introducing ten incorrect outcomes. We provide a full breakdown of the results in Appendix \ref{sec:more_results} (see Table \ref{tab:esbmc-progress}). \textcolor{black}{At he same time, the most recent results from 2026 demonstrated a performance drop with \esbmc{} correctly solving $114$ instances (compared to $162$ and $231$ in the two years before) and producing no incorrect verdicts.}

\section{Conclusions and Future Work}
\label{sec:future}

In this paper, we show how to model the floating-point details of neural networks by extracting their low-level implementation in plain C code. This step is necessary in ensuring the safety and reliability of neural network models, since the numerical details of performing finite-precision inference may affect the output of the computation and produce unwanted outcomes. Unfortunately, state-of-the-art neural network verifiers discard these computational details and thus are unable to guarantee the absence of vulnerabilities in the worst case.

For this reason, we perform a rigorous evaluation of the performance of another class of verification tools, i.e. software verifiers, which are able to reason on the exact behaviour of floating-point arithmetic. To do so, we significantly expand on \bench{}, our existing benchmark for neural network code verification. Our benchmark design satisfies two main requirements: the verdict of every verification query must be known in advance, and the set of verification queries must feature a gradual increase in complexity. Together, the two requirements allow us to test both the correctness and scalability of existing software verifiers.

In summary, our work reveals a number of crucial pieces of information. First, neural network verification at the software level is conceptually feasible, but the existing tools are not yet ready to cope with the scale of the problem. Indeed, $50\%$ of the software verifiers we test return a non-insignificant number of incorrect verdicts. Furthermore, even the most successful of them fail to solve a large majority of our benchmark. Second, we have evidence to show that the release of \bench{} has had a very beneficial impact on the community. In fact, the presence of our benchmark has led to a drastic reduction, if not complete elimination, of incorrect verdicts in most state-of-the-art software verification tools. By maintaining and expanding \bench{} over time, we commit to further encourage tool authors to improve their verifiers on floating-point tasks.

At the same time, we believe that the path towards certifying the whole implementation of a neural network against worst-case numerical deviations is still long. In particular, software verifiers need native support for mathematical libraries such as \texttt{math.h} and custom decision procedures for neural network code. In both cases, we plan to take inspiration from the algorithms employed by high-level neural network verifiers (branch and bound, linear abstractions) and adapt them to work on floating-point arithmetic.

\backmatter





\bmhead{Acknowledgements}

This work is partially funded by the EPSRC grant EP/T026995/1 entitled
``EnnCore: End-to-End Conceptual Guarding of Neural Architectures''
under \textit{Security for all in an AI enabled society}, the ARIA Opportunity Seed entitled ``Hardware-Level AI Safety Verification'' under \textit{Mathematics for Safe AI}, and the the EPSRC grant EP/V000497/1 entitled ``SCorCH: Secure Code for Capability Hardware''.










\begin{appendices}

\section{Adoption of the IEEE 754 Standard in the C Language}
\label{sec:ieee754_vs_c}

In Section \ref{sec:bench}, we present several instances of our benchmark where we rely on functions from the standard library \texttt{math.h}. There, we make the assumption that all functions in \texttt{math.h} have correctly-rounded implementations and existing software verifiers reason over correctly-rounded implementations. Crucially, if this assumption were not to hold, some of our ground-truth verdict may become incorrect. Here, we explain why this is not always the case because of some conflicts between the IEEE 754 floating-point and the C language semantics (see Sections \ref{sec:ieee_754} and \ref{sec:c99_math}). Note that the conflicts can be resolved by providing explicit operational models of the mathematical functions, as we propose in Section \ref{sec:op_models}.

\subsection{The IEEE 754 Standard}
\label{sec:ieee_754}

According to the IEEE 754 standard~\cite{IEEE754}, every floating-point operation should be correctly rounded. More specifically, Section 4.3 of the standard states the following:
\begin{quote}
    \textit{Except where stated otherwise, every operation shall be performed as if it first produced an intermediate result correct to infinite precision and with unbounded range, and then rounded that result according to one of the attributes in this clause.}
\end{quote}
At the same time, any programming language that conforms to the IEEE 754 standard is required to implement only a small subset of arithmetic operations. More specifically, Sections 5.4.1 and 5.5.1 of the standard lists the following mandatory operations: \textit{addition}, \textit{subtraction}, \textit{multiplication}, \textit{division}, \textit{square root}, \textit{fused multiply-add}, \textit{negation} and \textit{absolute value}. All other mathematical operations are labeled as ``recommended'', including \textit{transcendental}, \textit{hyperbolic} and \textit{trigonometric} functions (see Table 9.1 therein). That is, an IEEE 754 conformant programming language may not include functions such as the \textit{exponential} or \textit{sine} functions.

In summary, the IEEE 754 standard does not mandate the implementation of most functions we find the mathematical library \texttt{math.h}. However, if a programming language implements any of the recommended functions, then such an implementation must be correctly rounded. Indeed, Section 9.2 of the standard states that:
\begin{quote}
    \textit{A conforming operation shall return results correctly rounded for the applicable rounding direction for all operands in its domain.}
\end{quote}
Unfortunately, the IEEE 754 standard is not enforceable, and requires effort from programming language developers and users to maintain compliance. Section 11 of the standard recognises the potential danger of non-compliant languages to reproducibility, and states that:
\begin{quote}
    \textit{Reproducible results require cooperation from language standards, language processors, and users. A language standard should support reproducible programming.}
\end{quote}
Next, we show how the C language does not comply with the IEEE 754 standard.

\subsection{The C99 Standard}
\label{sec:c99_math}

The semantics of floating-point arithmetic have remained stable through the years. Here, we refer to the C99 standard~\cite{ISO:C99} as an example. Specifically, Section 5.2.4.2.2 of the standard states the following about the characteristics of floating point types in \texttt{float.h} (Point 5 therein):
\begin{quote}
    \textit{The accuracy of the floating-point operations (+, -, *, /) and of the library functions in \texttt{<math.h>} and \texttt{<complex.h>} that return floating-point results is \textbf{implementation defined}.}
\end{quote}
which is in direct constrast to the correctly-rounded requirement of the IEEE 754 standard (see Section \ref{sec:ieee_754}). Indeed, the only mention of correct rounding behaviour regards the \textit{fused multiply-add} operator.\footnote{Contrary to popular belief, the C99 standard does not explicitly mandate correctly-rounded behaviour for the square root function (see Section 7.12.7.5 therein).} Specifically, Section 7.12.13.1 of the C99 standard states that:
\begin{quote}
    \textit{The \texttt{fma} functions compute $(x \times y) + z$, rounded as one ternary operation: they compute the value (as if) to infinite precision and round once to the result format, according to the rounding mode characterized by the value of \texttt{FLT\_ROUNDS}. A range error may occur.}
\end{quote}
In summary, the C99 standard provides an ambiguous specification of the semantics of floating-point arithmetic in the C language.\footnote{The newest C23 standard makes an effort to make \texttt{math.h} fully compliant with IEEE 754, see \url{https://www.open-std.org/jtc1/sc22/wg14/www/docs/n3220.pdf}.} This fact severely hinders the testing and verification of numerical software, which requires a clear and non-ambiguous specification~\cite{Cody1982floating,Brisebarre2024correctly}. Indeed, existing verification tools cope with the inherent ambiguity in the C language by introducing various degrees of non-determinism and numerical approximation in their operational models of \texttt{math.h} functions. We show the consequences of this choice in the experimental results of Section \ref{sec:results_maths}.

\section{Differences between \bench{1.0} and \bench{2.0}}
\label{sec:version_diff}

In this appendix, we list all differences between the earliest release of \bench{1.0} as described in our technical report~\cite{manino2023neurocodebench} and the current release of \bench{2.0}.

\paragraph{Bug Fixes.} During benchmark preparation for the 2024 edition of SV-COMP, the community discovered a few minor bugs in \bench{1.0}. We list them here, with the corresponding patches:
\begin{itemize}
    \item \texttt{lunarlander\_26\_unsafe.c} and \texttt{lunarlander\_90\_safe.c} contained wrong pre-conditions due to a missing exponential notation in the constant bounds.\footnote{\url{https://gitlab.com/sosy-lab/benchmarking/sv-benchmarks/-/merge\_requests/1490}} We reintroduced the exponential notation, thus making the pre-condition non-empty.
    \item \texttt{cos\_7\_safe.c}, \texttt{erf\_5\_safe.c}, \texttt{logistic\_7\_unsafe.c}, \texttt{softsign\_5\_safe.c} and \texttt{tanh\_7\_safe.c} contained calls to the \texttt{isgreaterequal} macro with an integer argument.\footnote{\url{https://gitlab.com/sosy-lab/benchmarking/sv-benchmarks/-/merge\_requests/1456}} We replaced the arguments with a floating-point constant to match the macro type.
    \item \texttt{logistic\_1\_safe.c} and \texttt{tanh\_1\_safe.c} contained post-conditions with the wrong inequality sign.\footnote{\url{https://gitlab.com/sosy-lab/benchmarking/sv-benchmarks/-/issues/1394}} We replaced the erroneous call to \texttt{isgreaterequal} with \texttt{islessequal}, which solved the issue.
\end{itemize}
Furthermore, we independently discovered the following additional bug, which was patched in time for SV-COMP 2024:
\begin{itemize}
    \item \texttt{tanh\_w8\_r4\_case\_1\_safe.c} in the Hopfield network subcategory had the wrong verdict due to the output converging to $1$, but not quite reaching it for all inputs. We renamed it to \texttt{tanh\_w8\_r4\_case\_1\_unsafe.c} accordingly.
\end{itemize}

\paragraph{New Instances.} Version 2.0 of our benchmark introduces $305$ new verification queries, which we have clearly marked with the tag \labelnew{} throughout the main body of the paper. For ease of reference, we list them here as well:
\begin{itemize}
    \item \textit{Maths Functions.} We introduce $14$ new instances, covering properties of the \texttt{expm1f} and \texttt{log1pf} functions.
    \item \textit{Activation Functions.} We introduce $15$ new instances, covering the Gaussian, GLU, and Step activations.
    \item \textit{Neural Layers.} The $86$ instances in this category are almost entirely new, with the exception of the $14$ properties of the SoftMax layer. The latter were originally part of the activation function category in \bench{1.0}.
    \item \textit{SAT ReLU Networks.} This category is entirely new, for a total of $96$ instances.
    \item \textit{Lipschitz-Bounded Networks.} This category is entirely new, for a total of $108$ instances.
\end{itemize}

\paragraph{Operational Models.} Version 2.0 of our benchmark can be augmented with operational models of \texttt{math.h} functions. We discuss their role in depth throughout the paper (see Sections \ref{sec:setup_maths}, \ref{sec:op_models} and \ref{sec:workflow}). Here, we list the single-precision floating-point functions they provide:
\begin{itemize}
    \item \textit{\musl{}}. Provides \texttt{acosf}, \texttt{asinf}, \texttt{atanhf}, \texttt{cosf}, \texttt{erff}, \texttt{expf}, \texttt{expm1f}, \texttt{fabsf}, \texttt{fmaxf}, \texttt{log1pf}, \texttt{logf}, \texttt{sinf}, and \texttt{tanhf}. The addition of \texttt{sqrtf} is optional.
    \item \textit{\coremath{}}. Provides \texttt{acosf}, \texttt{asinf}, \texttt{atanhf}, \texttt{cosf}, \texttt{erff}, \texttt{expf}, \texttt{expm1f}, \texttt{log1pf}, \texttt{logf}, \texttt{sinf}, and \texttt{tanhf}.
\end{itemize}

\section{Observations on the Behaviour of Verifiers on \bench{2.0}}
\label{sec:observations}

\begin{figure}[t]
    \centering
    \includegraphics[width=\linewidth]{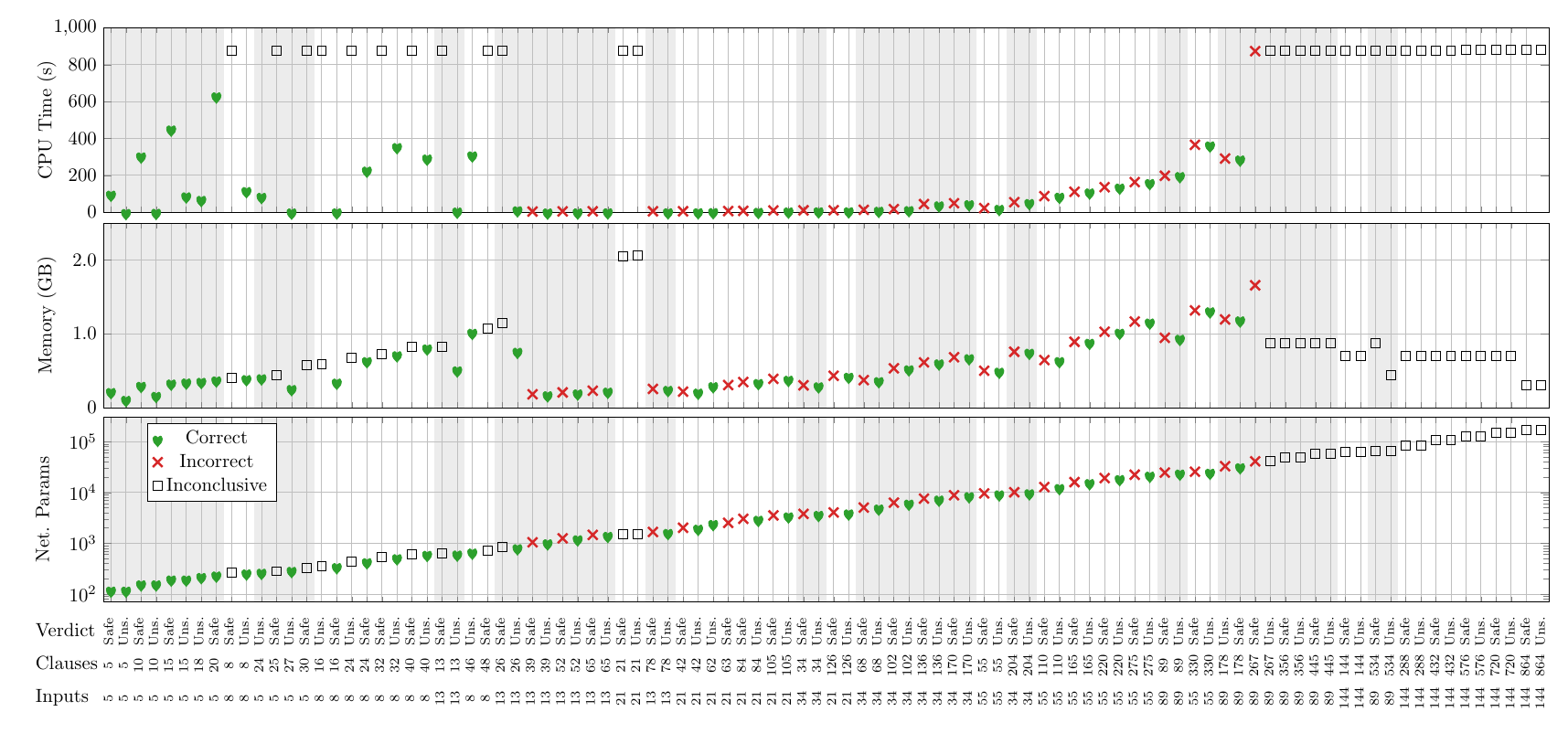}
    \caption{Performance of \cbmc{} on the SAT ReLU Networks category.}
    \label{fig:sat_relu_cbmc}
\end{figure}

\begin{figure}[t]
\centering
    \begin{subfigure}[b]{0.49\textwidth}
        \includegraphics[width=\textwidth]{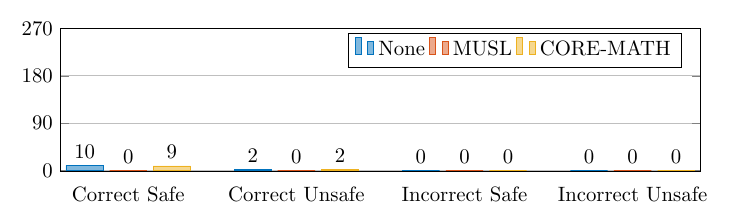}
        \caption{\twols{}}
        \label{fig:twols_no_math}
    \end{subfigure}
    \hfill
    \begin{subfigure}[b]{0.49\textwidth}
        \includegraphics[width=\textwidth]{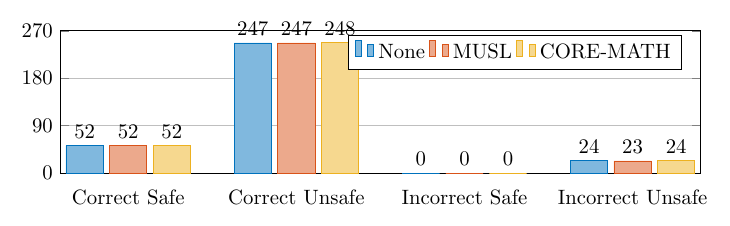}
        \caption{\cbmc{}}
        \label{fig:cbmc_no_math}
    \end{subfigure}
    \hfill\begin{subfigure}[b]{0.49\textwidth}
        \includegraphics[width=\textwidth]{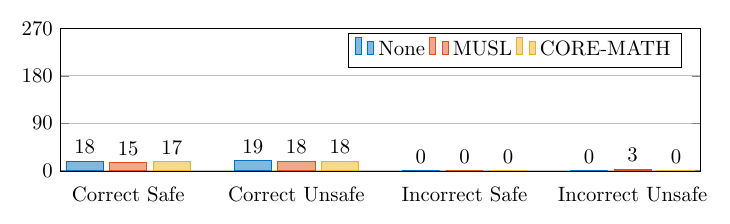}
        \caption{\cpachecker{}}
        \label{fig:cpa_no_math}
    \end{subfigure}
    \hfill\begin{subfigure}[b]{0.49\textwidth}
        \includegraphics[width=\textwidth]{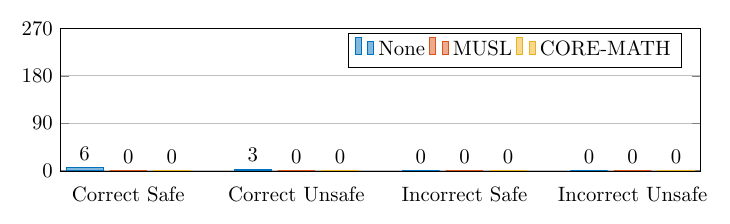}
        \caption{\divine{}}
        \label{fig:divine_no_math}
    \end{subfigure}
    \hfill\begin{subfigure}[b]{0.49\textwidth}
        \includegraphics[width=\textwidth]{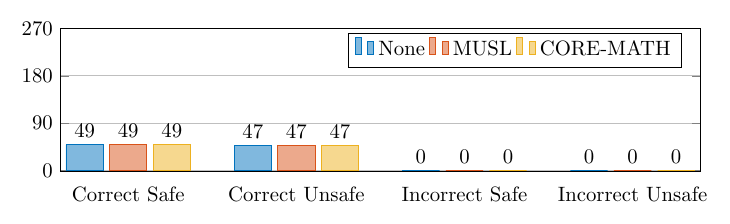}
        \caption{\esbmc{}}
        \label{fig:esbmc_no_math}
    \end{subfigure}
    \hfill\begin{subfigure}[b]{0.49\textwidth}
        \includegraphics[width=\textwidth]{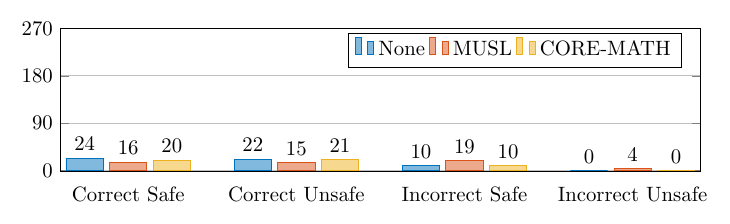}
        \caption{\pesco{}}
        \label{fig:pesco_no_math}
    \end{subfigure}
    \hfill\begin{subfigure}[b]{0.49\textwidth}
        \includegraphics[width=\textwidth]{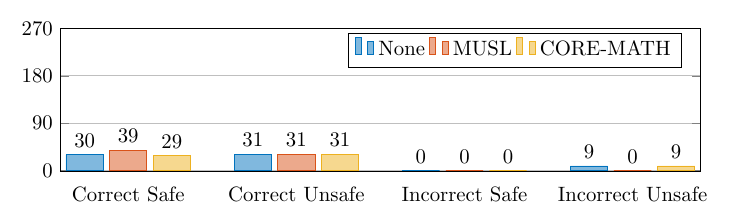}
        \caption{\pinaka{}}
        \label{fig:pinaka_no_math}
    \end{subfigure}
    \hfill\begin{subfigure}[b]{0.49\textwidth}
        \includegraphics[width=\textwidth]{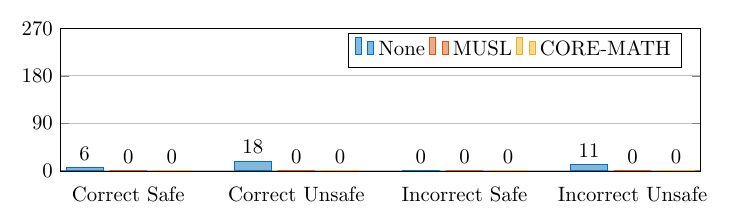}
        \caption{\uautomizer{}}
        \label{fig:uauto_no_math}
    \end{subfigure}
\caption{Verifier performance on the $732$ benchmark instances of \bench{2.0} without \texttt{math.h} functions.}
\label{fig:no-math-funs}
\end{figure}

In this appendix, we report our detailed observation on the behaviour of the software verifiers we tested in Section \ref{sec:exp}. As we take the perspective of \textit{expert users} in the course of our experiments, the content of this appendix should not be seen as an attempt to discover the root causes of the behaviour we observe. Instead, we report them here to provide additional context around our experimental results.
\begin{itemize}
    \item \textit{\twols{}}. This verifier clearly applies an explicit heuristic to detect maths functions. The heuristic quickly causes \twols{} to return ``unknown'' as verdict. The output logs reveal the specific list of functions:
    \begin{quote}
        \texttt{2ls-binary: preprocessing\_util.cpp:701: \\void twols\_parse\_optionst::assert\_no\_unsupported\_function\_calls(goto\_modelt\&): Assertion `NOT\_MATH\_FUN(name, "cos") \&\& NOT\_MATH\_FUN(name, "acos") \&\& NOT\_MATH\_FUN(name, "sin") \&\& NOT\_MATH\_FUN(name, "asin") \&\& NOT\_MATH\_FUN(name, "tan") \&\& NOT\_MATH\_FUN(name, "atan") \&\& NOT\_MATH\_FUN(name, "atan2") \&\& NOT\_MATH\_FUN(name, "cosh") \&\& NOT\_MATH\_FUN(name, "acosh") \&\& NOT\_MATH\_FUN(name, "sinh") \&\& NOT\_MATH\_FUN(name, "asinh") \&\& NOT\_MATH\_FUN(name, "tanh") \&\& NOT\_MATH\_FUN(name, "atanh") \&\& NOT\_MATH\_FUN(name, "exp") \&\& NOT\_MATH\_FUN(name, "exp2") \&\& NOT\_MATH\_FUN(name, "expm1") \&\& NOT\_MATH\_FUN(name, "log") \&\& NOT\_MATH\_FUN(name, "log10") \&\& NOT\_MATH\_FUN(name, "log2") \&\& NOT\_MATH\_FUN(name, "log1p") \&\& NOT\_MATH\_FUN(name, "erf")' failed.}
    \end{quote}
    \item \textit{\cbmc{}}. This verifier returns $121$ incorrect unsafe results on \bench{2.0} as shown in Table \ref{tab:plain-verifiers-all-verdicts}. In contrast to all other verifiers except \esbmc{}, the tool produces a full counterexample trace for each unsafe verdict. By manual inspection, these traces tend to include zero and NaN values as inputs. Moreover, a large majority of incorrect results seems to be due to under-specified maths functions: the return value of most calls to \texttt{math.h} is a non-deterministic float, sometime restricted to a reasonable range (e.g. the output of \texttt{cosf} is restricted to $[-1,1]$). At the same time, \cbmc{} also returns incorrect unsafe verdicts for benchmarks that do \textit{not} contain calls to \texttt{math.h}, as shown in Figure \ref{fig:sat_relu_cbmc}. Reproducing the corresponding counterexample traces yield output values that do \textit{not} violate the network post-condition, thus confirming that they are spurious.
    \item \textit{\cpachecker{}}. This verifier reports an error when processing some functions from the C standard. For instance, verification queries that contain calls to \texttt{memset} trigger the following log message:
    \begin{quote}
        \texttt{Error: line 11207: Unrecognized C code (Memory assignment function called but their handling is disabled. Set cpa.predicate.enableMemoryAssignmentFunctions=true to enable.): memset(C, 0, (outrows * outcols) * 4UL) (CExpressionVisitorWithPointerAliasing.visit, SEVERE)}
    \end{quote}
    Similar error messages are returned for most maths functions, even when \musl{} or \coremath{} operational models are provided. For example, benchmark instances that rely on \texttt{tanhf} cause the following error:
    \begin{quote}
        \texttt{Error: line 2154: Unsupported feature (arithmetic function): tanhf(x) (line was originally float y = tanhf(x);) (ExpressionToFormulaVisitor.visit, SEVERE)}
    \end{quote}
    \item \textit{\divine{}}. Unfortunately, this verifier provides no information in the log files.
    \item \textit{\esbmc{}}. This verifier internally uses a subset of the \musl{} operational models~\cite{Menezes2024esbmc}, as we mention in Section \ref{sec:op_models}. Interestingly, explicitly including the \musl{} models as part of the benchmark causes $3$ incorrect unsafe verdicts to appear (see Figure \ref{fig:esbmc_math}). Note that instances that do not contain maths functions do not cause this issue, as shown in Figure \ref{fig:esbmc_no_math}. Since \esbmc{} produces full counterexample traces for its unsafe verdicts, we can run the three benchmark instances with the given concrete inputs. Doing so leads to outputs that do \textit{not} violate the assertion, thus showing that the counterexamples are indeed spurious.
    \item \textit{\pesco{}}. This verifier throws the same errors as CPAChecker, except for the \texttt{memset} ones. For example, the below is a snippet taken verbatim from the output logs:
    \begin{quote}
        \texttt{Error: line 1129: Unsupported feature (arithmetic function): tanhf(x1) (line was originally float y1 = tanhf(x1);) (ExpressionToFormulaVisitor.visit, SEVERE)}
    \end{quote}
    \item \textit{\pinaka{}}. This verifier returns $82$ incorrect unsafe results (see Table \ref{tab:plain-verifiers-all-verdicts}), but does not provide counterexample traces that we can inspect. Furthermore, the logs do not contain errors or warnings regarding the absence maths function support.
    \item \textit{\uautomizer{}}. This verifier appears to process each benchmark instance in two stages: default analysis and bit-precise analysis. The first stage often fails due to the presence of basic floating point operations such as \texttt{fabs}, \texttt{fmax}, \texttt{isnan} and \texttt{round}. The following is an example from the output logs:
    \begin{quote}
          \texttt{ExceptionOrErrorResult: UnsupportedOperationException: floating point operation not supported in non-bitprecise translation: fabs}
    \end{quote}
    Then, the verifier tries again in bit-precise mode. From the phrasing of the log messages, we can infer that this mode must still involve some form of over-approximation. For example, a benchmark instance containing calls to \texttt{tanhf} may cause the following message:
    \begin{quote}
        \texttt{Unable to prove that a call to reach\_error is unreachable. Reason: overapproximation of tanhf at line 1129.}
    \end{quote}
    When adding \musl{} or \coremath{} operational models, the analysis reaches the bit-precise stage but fails due to unsupported functions. This happens even for benchmarks where all mathematical operations could be sliced away, e.g. those containing only ReLU activation functions.
\end{itemize}

For completeness, we report a full breakdown of the verifiers behaviour in Table \ref{tab:verifiers-all-verdicts}.

\begin{table}[ht!]
\caption{Detailed comparison of the verdicts returned by the verifiers on \bench{2.0} using 3 different operational model settings: None, \musl{}, and \coremath{}.}
\label{tab:verifiers-all-verdicts}
    \centering
    \begin{tabular}{c|c|cc|cc|cccc}
        \toprule
        \multirow{2}{*}{Verifier} & 
        \multirow{2}{*}{Op. Models} & 
        \multicolumn{2}{c|}{Correct} & 
        \multicolumn{2}{c|}{Incorrect} & 
        \multicolumn{4}{c}{Inconclusive}\\
        \cmidrule{3-10}
                                & & True & False & True & False & Timeout & Out of Memory & Unknown & Error \\
        \midrule
        \multirow{3}{*}{\twols}           & None & 12 & 2 & - & - & 4 & - & 894 & -\\
                                                & \musl{} & - & - & - & - & - & - & 912 & -\\
                                                & \coremath{} & 11 & 2 & - & - & 5 & - & 894 & -\\
        \midrule
        \multirow{3}{*}{\cbmc}          & None & 66 & 305 & - & 121 & - & 80 & - & 340\\
                                                & \musl{} & 75 & 279 & - & 25 & - & 90 & 30 & 413\\
                                                & \coremath{} & 76 & 299 & - & 69 & - & 97 & 11 & 360\\
        \midrule
        \multirow{3}{*}{\cpachecker}    & None & 18 & 19 & - & - & 292 & 265 & - & 318\\
                                                & \musl{} & 18 & 28 & - & 17 & 392 & 264 & - & 193\\
                                                & \coremath{} & 18 & 21 & - & - & 335 & 265 & - & 273\\
        \midrule
        \multirow{3}{*}{\divine{}}        & None & 6 & 3 & - & - & 390 & 152 & 361 & -\\
                                                & \musl{} & - & - & - & - & - & - & 912 & -\\
                                                & \coremath{} & - & - & - & - & - & - & 912 & -\\
        \midrule
        \multirow{3}{*}{\esbmc{}}         & None & 76 & 86 & - & - & 539 & 211 & - & -\\
                                                & \musl{} & 67 & 80 & - & 3 & 551 & 211 & - & -\\
                                                & \coremath{} & 63 & 68 & - & - & 570 & 211 & - & -\\
        \midrule
        \multirow{3}{*}{\pesco{}}         & None & 24 & 25 & 10 & - & 482 & 58 & - & 313\\
                                                & \musl{} & 19 & 27 & 27 & 18 & 598 & 54 & - & 169\\
                                                & \coremath{} & 21 & 27 & 10 & - & 541 & 60 & - & 253\\
        \midrule
        \multirow{3}{*}{\pinaka{}}        & None & 44 & 71 & - & 82 & 243 & 292 & - & 180\\
                                                & \musl{} & 59 & 64 & - & 5 & 305 & 293 & - & 186\\
                                                & \coremath{} & 53 & 59 & - & 24 & 302 & 292 & - & 182\\
        \midrule
        \multirow{3}{*}{\uautomizer{}}    & None & 9 & 27 & - & 12 & 754 & 1 & 106 & 3\\
                                                & \musl{} & - & - & - & - & - & - & 912 & -\\
                                                & \coremath{} & - & - & - & - & - & - & 912 & -\\
        \bottomrule
    \end{tabular}
\end{table}

\section{Time Distribution Graphs}
\label{sec:time_distro}

\begin{figure}[t]
    \centering
    \includegraphics[width=0.75\linewidth]{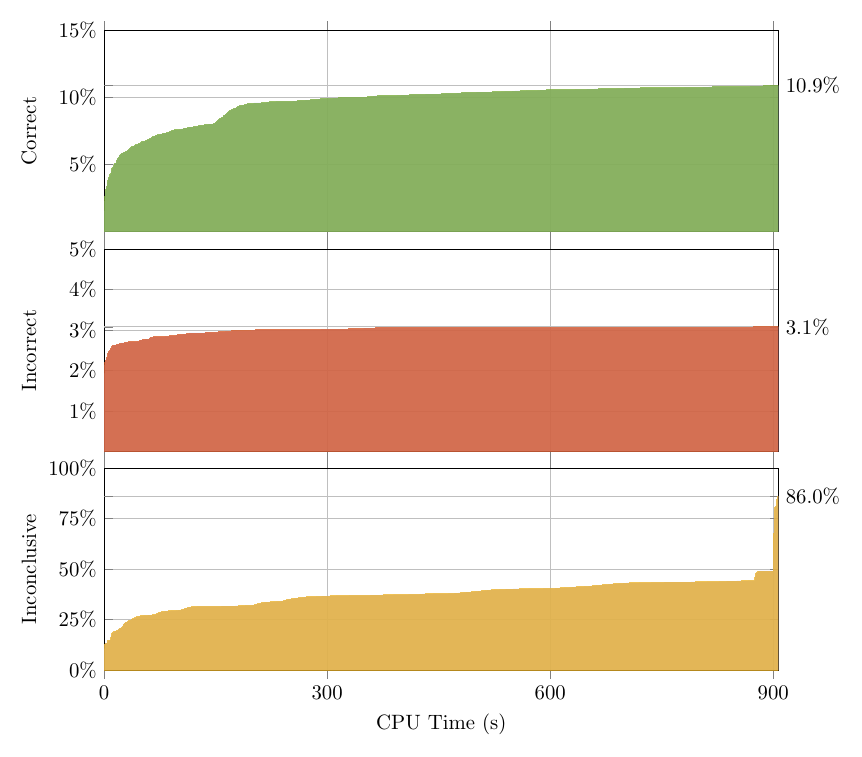}
    \caption{Cumulative distribution of the time taken by all verifiers to solve the \bench{2.0} benchmark. On the right, we report the overall percentage of correct, incorrect, and inconclusive verdicts.}
    \label{fig:all_timings}
\end{figure}

In Figure \ref{fig:all_timings}, we report the cumulative distribution of the time it takes the verifiers to produce a verdict. Specifically, we compare the distributions of correct, incorrect, and inconclusive verdicts. The results show that almost the entirety of incorrect verdicts (either safe or unsafe) are returned in the first few seconds of computation. In contrast, correct verdicts are more uniformly spread across the $900$ seconds of computation time. As expected, more than half of the inconclusive verdicts are due to timeout, i.e. they are close to the $900$ second threshold. Still, almost a third happen in the first seconds of computation, thus reinforcing our conclusion that many verifiers simply ``reject'' a large number of benchmark instances (see also Appendix \ref{sec:observations}).

\section{Additional Information}
\label{sec:more_results}

\begin{table}[t]
\caption{Detailed comparison of different versions of \esbmc{} on \bench{2.0}, spanning the years 2018 to 2025.}
\label{tab:esbmc-progress}
    \centering
    \begin{tabular}{c|cc|cc|cccc}
        \toprule
        \multirow{2}{*}{Year} & 
        \multicolumn{2}{c|}{Correct} & 
        \multicolumn{2}{c|}{Incorrect} & 
        \multicolumn{4}{c}{Inconclusive}\\
        \cline{2-9}
                                & True & False & True & False & Timeout & Out of Memory & Unknown & Error \\
        \midrule
        2018 & 83 & 27 & 10 & 73 & 301 & 342 & 76 & - \\
        2019 & 113 & 57 & 34 & 89 & 393 & 226 & - & - \\
        2020 & 114 & 52 & 34 & 89 & 394 & 227 & 2 & - \\
        2021 & 119 & 92 & 34 & 81 & 359 & 227 & - & - \\
        2022 & 350 & 71 & 234 & 81 & 159 & 17 & - & - \\
        2023 & 345 & 91 & 210 & 85 & 69 & 16 & 96 & - \\
        2024 & 76 & 86 & - & - & 539 & 211 & - & - \\
        2025 & 134 & 97 & 10 & - & 459 & 212 & - & - \\
        2026 & 59 & 55 & - & - & 798 & - & - & - \\
        \bottomrule
    \end{tabular}
\end{table}

This Appendix contains additional information that would not fit the main text of the present paper:
\begin{itemize}
    \item \textit{Historical Progress.} Table \ref{tab:esbmc-progress} lists the data behind the historical comparison of different \esbmc{} versions in Figure \ref{fig:esbmc_history}
    \item \textit{Math Library Support.} Table \ref{sec:results_maths} reports the full list of functions from \texttt{math.h} that we use when filtering the relevant subset of \bench{2.0} in Section \ref{sec:results_maths}. Note that the table is conservative, i.e. it contains functions that \bench{2.0} instances may not call, at least directly.
\end{itemize}

\begin{table}[t]
\caption{Full list of functions that we use to filter the $180$ instances of \bench{2.0} in Section \ref{sec:results_maths}.}
\label{tab:math-h-functions}
    \centering
    \begin{tabular}{l|l}
    \toprule
    \multirow{2}{*}{Exponential functions}     & \texttt{exp},\texttt{expf},\texttt{expl},\texttt{exp2},\texttt{exp2f},\texttt{exp2l},\texttt{expm1},\texttt{expm1f},\texttt{expm1l}, \\
    & \texttt{log},\texttt{logf},\texttt{logl},\texttt{log10},\texttt{log10f},\texttt{log10l},\texttt{log2},\texttt{log2f},\texttt{log2l},\texttt{log1p},\texttt{log1pf},\texttt{log1pl}  \\
    \midrule
    Power functions     & \texttt{pow},\texttt{powf},\texttt{powl}, \texttt{sqrt},\texttt{sqrtf},\texttt{sqrtl}, \texttt{cbrt},\texttt{cbrtf},\texttt{cbrtl},\texttt{hypot},\texttt{hypotf},\texttt{hypotl} \\
    \midrule
    \multirow{2}{*}{Trigonometric functions} & \texttt{sin},\texttt{sinf},\texttt{sinl},\texttt{cos},\texttt{cosf},\texttt{cosl},\texttt{tan},\texttt{tanf},\texttt{tanl},\\
    
    & \texttt{asin},\texttt{asinf},\texttt{asinl},\texttt{acos},\texttt{acosf},\texttt{acosl},\texttt{atan},\texttt{atanf},\texttt{atanl},\texttt{atan2},\texttt{atan2f},\texttt{atan2l} \\
    \midrule
    \multirow{2}{*}{Hyperbolic functions} & \texttt{sinh},\texttt{sinhf},\texttt{sinhl},\texttt{cosh},\texttt{coshf},\texttt{coshl},\texttt{tanh},\texttt{tanhf},\texttt{tanhl},\\
    
    & \texttt{asinh},\texttt{asinhf},\texttt{asinhl},\texttt{acosh},\texttt{acoshf},\texttt{acoshl},\texttt{atanh},\texttt{atanhf},\texttt{atanhl} \\
    \midrule
    Error functions & \texttt{erf},\texttt{erff},\texttt{erfl},\texttt{erfc},\texttt{erfcf},\texttt{erfcl} \\
    \midrule
    Gamma functions & \texttt{tgamma},\texttt{tgammaf},\texttt{tgammal},\texttt{lgamma},\texttt{lgammaf},\texttt{lgammal}\\
    \midrule
    \multirow{3}{*}{Floating-point manipulation functions} & \texttt{frexp},\texttt{frexpf},\texttt{frexpl},\texttt{ldexp},\texttt{ldexpf},\texttt{ldexpl}, \\
    
    & \texttt{scalbn},\texttt{scalbnf},\texttt{scalbnl},\texttt{scalbln},\texttt{scalblnf},\texttt{scalblnl}, \\
    
    & \texttt{ilogb},\texttt{ilogbf},\texttt{ilogbl},\texttt{logb},\texttt{logbf},\texttt{logbl} \\
    \bottomrule
    \end{tabular}
\end{table}




\end{appendices}


\bibliography{ref_v3.bib}

\end{document}